\documentclass[12pt,preprint]{aastex}
\newcommand{\nraoblurb}{The National Radio Astronomy Observatory is
a facility of the National Science Foundation operated under cooperative
agreement by Associated Universities, Inc.}

\newcommand{\cm}{$\,{\rm cm}$}
\newcommand{\kpc}{$\,{\rm kpc}$}
\newcommand{\K}{\ensuremath{\,{\rm K}}}

\newcommand{\ghz}{$\,{\rm GHz}$}
\newcommand{\percc}{$\,{\rm cm^{-3}}$}

\newcommand{\kms}{\ensuremath{\,{\rm km\,s}^{-1}}}

\newcommand{\jy}{\ensuremath{{\rm \,Jy}}}
\newcommand{\jyb}{$\rm \,Jy\,beam^{-1}$}

\newcommand{\hi}{{\rm H\,}{{\sc i}}}
\newcommand{\hii}{{\rm H\,}{{\sc ii}}}

\newcommand{\co}{\ensuremath{^{12}{\rm CO}}}
\newcommand{\cor}{\ensuremath{^{13}{\rm CO}}}
\newcommand{\wco}{\ensuremath{W_{\rm CO}}}
\newcommand{\nexpo}[2]{\ensuremath{{#1}\times 10^{#2}}}
\newcommand{\expo}[1]{\ensuremath{10^{#1}}}

\newcommand{\tex}{$T_{\rm ex}$}
\slugcomment{For ApJSuppl: final draft --- lda, \today}

\shorttitle{CO in Galactic H~{\small \bf II} Regions}
\shortauthors{Anderson et al.}

\begin{document}

\title{The Molecular Properties of Galactic H\,{\small \bf II} Regions}

\author{L. D. Anderson\altaffilmark{1}, T. M. Bania\altaffilmark{1}, 
J. M. Jackson\altaffilmark{1}, D. P. Clemens\altaffilmark{1}, 
M. Heyer\altaffilmark{2}, R. Simon\altaffilmark{3}, R. Y. Shah\altaffilmark{4},
\& J. M. Rathborne\altaffilmark{5}}

\altaffiltext{1}{Institute for Astrophysical Research,
725 Commonwealth Ave., Boston University, Boston MA 02215, USA.}
\altaffiltext{2}{Department of Astronomy, University of Massachusetts,
Amherst, MA 01003, USA.}
\altaffiltext{3}{Physikalisches Institut, Universit\"{a}t zu K\"{o}ln,
50937 Cologne, Germany}
\altaffiltext{4}{MIT Lincoln Laboratory, 244 Wood Street, Lexington, MA 02420}
\altaffiltext{5}{Harvard-Smithsonian Center for Astrophysics, 60 Garden
St., Cambridge, MA 02138, USA}

\begin{abstract}

We derive the molecular properties for a sample of 301 Galactic \hii\
regions including 123 ultra compact (UC), 105 compact, and 73 diffuse
nebulae. We analyze all sources within the BU-FCRAO Galactic Ring
Survey (GRS) of \cor\ emission known to be \hii\ regions based upon
the presence of radio continuum and \cm-wavelength radio recombination
line emission.  Unlike all previous large area coverage \cor\ surveys,
the GRS is fully sampled in angle and yet covers $\sim 75$ square
degrees of the Inner Galaxy.  The angular resolution of the GRS
(46\arcsec ) allows us to associate molecular gas with \hii\ regions
without ambiguity and to investigate the physical properties of this
molecular gas.  We find clear CO/\hii\ morphological associations in
position and velocity for $\sim 80\%$ of the nebular sample.  Compact
\hii\ region molecular gas clouds are on average larger than UC
clouds: $2 \farcm 2$ compared to $1 \farcm 7$.  Compact and UC \hii\
regions have very similar molecular properties, with $\sim 5 ~{\rm K}$
line intensities and $\sim 4 \kms$ line widths.  The diffuse \hii\
region molecular gas has lower line intensities, $\sim 3 ~{\rm K}$,
and smaller line widths, $\sim 3.5 \kms$.  These latter characteristics
are similar to those found for quiescent molecular clouds in the GRS.
Our sample nebulae thus show evidence for an evolutionary sequence
wherein small, dense molecular gas clumps associated with UC \hii\
regions grow into older compact nebulae and finally fragment and
dissipate into large, diffuse nebulae.

\end{abstract}


\section{INTRODUCTION\label{sec:intro}}

In the classical view of \hii\ region formation OB stars form inside Giant
Molecular Clouds (GMCs).  Any newly formed OB star emits extreme
ultraviolet (EUV) radiation ($> 13.6$ eV) and ionizes the surrounding
medium of a molecular cloud, creating an \hii\ region. The ionizing
photons have more energy than is necessary to ionize the gas, however,
and thus some excess energy heats the ambient gas. Because of the
pressure difference between the cold, natal molecular cloud ($\sim$\,30
K) and the \hii\ region ($\sim$\,\expo{4}\,K), the ionization front
expands into the molecular cloud.

A photodissociation region (PDR) exists beyond the ionization
front. The PDR is a zone where photons of energies lower than 13.6 eV
ionize elements with low ionization potentials and dissociate
molecules \citep[see][ for a review of dense PDRs]{hollenbach97}.
Within the PDR there are three important boundaries. At $A_V$ $\simeq
1$, a dissociation front exists where $\rm H_2$ replaces H as the
dominant species. Further from the emitting star, at $A_V \simeq 4$,
there is a boundary between $\rm C^{+}/ \rm C/ \rm CO$ where carbon
becomes stored predominantly in the CO molecule. The final boundary of
the PDR occurs at the $\rm O/ \rm O_2$ transition, $A_V \simeq
10$. The PDR is therefore a transition region between the \hii\ region
and the molecular cloud; it is completely ionized at one boundary and
completely molecular at the other. Since all \hii\ regions should
produce PDRs, in this simplified view all \hii\ regions should have
associated molecular gas when they first form.  As the nebulae evolve,
however, the OB stars can travel far enough to leave the environs of
their natal molecular clouds.

Complicating this simple scenario is the fact that molecular clouds are
clumpy and inhomogeneous on all scales \citep[e.g.,][]{falgorone96,
kramer98, simon01}.  In a clumpy medium, EUV photons can penetrate to
different depths, creating non-spherical ionization
fronts. Additionally, \hii\ regions evolve by moving away from their
natal cloud environments and displacing the local gas. Understanding the
complicated interaction between young stars, \hii\ regions, and
molecular gas is crucial to the study of massive star formation and the
impact massive stars have on their environment.

The molecular component of \hii\ regions has been studied in detail by
many authors. Most studies, however, have focused mainly on
ultra-compact (UC) \hii\ regions \citep[e.g.,][]{churchwell90, kim03}.
When compact \hii\ regions were observed in CO, they were often only
observed with single pointings using single-dish telescopes
\citep[e.g.,][]{brand84, whiteoak82, russeil04}. As our results show,
CO gas is frequently offset from the nominal position of an \hii\
region. A completely sampled map is required to understand fully the
dynamics and properties of molecular gas associated with \hii\
regions. Furthermore, a sample with \hii\ regions at all evolutionary
stages is necessary to understand how this interaction progresses as
the \hii\ region ages.

Here we describe a large scale study of the molecular properties of
diffuse, compact, and ultra-compact Galactic \hii\ regions using fully
sampled CO maps.  We trace these molecular properties using the \cor\,
$J= 1\rightarrow 0$ emission mapped by the Boston University--Five
College Radio Astronomy Observatory (hereafter BU--FCRAO) Galactic
Ring Survey
\citep[GRS: ][]{jackson06}.  CO is an excellent tracer of molecular
material because of its high abundance and the fact that it is
rotationally excited at densities common in molecular clouds
($\ga$\,500\percc).

\section{Milky Way Surveys\label{surveys}}

\subsection{Galactic H~{\bf\small II} Regions\label{catalog}}

Our Galactic \hii\ region source sample is compiled from the
\hii\ region radio recombination line (RRL) catalog of \citet{lockman89}
(hereafter L89), which consists of nearly 500 RRL observations with a
3\arcmin\ beam at positions of known continuum emission. These continuum
emission sources were drawn from the 5 \ghz\, continuum survey of
\citet{altenhoff79}.  All positions in \citet{altenhoff79} with a peak
flux density greater than 1 \jyb\ and many sources with peak flux
densities down to 0.8 \jyb\ were observed, unless the source was in a
very confused region or was a known supernova remnant (SNR).
Since \citet{altenhoff79} has coverage from $l=357\fdg 5$ to $60\degr$
and $\vert b\vert < 1\degr$, the L89 catalog is complete down to a flux
limit of at least 1 \jy\ at 5 \ghz\ in this region of the Galaxy. This
sky coverage entirely overlaps and is much larger than the GRS survey
zone (discussed in \S \ref{GRS}), so the L89 catalog is complete at
least down to 1\jyb\ over the extent of the GRS.
The sources in L89 are the classical \hii\ regions first detected in
the 1970's at cm-wavelengths using RRLs.  Although they are often
misidentified, these are not the ultra compact, high density \hii\
regions that are being studied mostly using radio interferometers.
Below, to distinguish these classical \hii\ regions from UC nebulae,
we will call the L89 nebulae ``compact'' \hii\ regions.

Our nebular sample also includes diffuse \hii\ regions from
\citet{lockman96} (hereafter L96). This catalog consists of RRL
measurements at 6\cm\ ($\sim 6\arcmin$ beam) and 9\cm\ ($\sim 9\arcmin$
beam) toward 130 faint, extended continuum sources.  Most of these
sources are drawn from \citet{altenhoff79} and have peak flux densities
greater than 0.5 \jyb.  The L89 survey is actually the pilot study of
this diffuse sample; it contains 40 diffuse sources.

Recently a large catalog of 1442 Galactic \hii\ regions was compiled
from 24 published studies of Galactic \hii\ regions
\citep{paladini03}. These \hii\ regions were found using single-dish,
medium resolution (few arcminute beamwidths) observations. For our
purposes, however, this catalog is not useful because nebular positions
are given to an accuracy of only 6\arcmin.

Finally, our source sample contains UC \hii\ regions taken from
\citet{wc89a}, \citet{kurtz94}, \citet{watson03}, and
\citet{sewilo04}. The \citet{wc89a} UC sources were selected by: (1) the
presence of a small or unresolved radio source, (2) a spectrum
consistent with free-free emission, and (3) strong FIR emission. The UC
sources from \citet{kurtz94}, \citet{watson03}, and \citet{sewilo04}
were selected based on the \citet{wc89b} criteria: {\it IRAS} flux
density ratios log$(F_{25}/F_{12}) \ge 0.57$, log$(F_{60}/F_{12}) \ge
1.30$, and $F_{100} \ge 1000 \jy$ ($\ge 700 \jy$ for \citet{watson03}),
where $F_{\lambda}$ is the {\it IRAS} flux density at $\lambda\,
\micron$.
\subsection{{\bf The $^{\bf 13}$}CO Galactic Ring Survey\label{GRS}}

We use the BU--FCRAO \cor\, Galactic Ring Survey data\footnote[1]{Data
available at http://www.bu.edu/galacticring/} \citep{jackson06} to
characterize the molecular properties of all \hii\ regions in the GRS. The GRS traces the 5 \kpc\ molecular ring discovered by
\citet{burton75} and \citet{ss75}.  This annulus of enhanced CO emission
dominates the inner Galaxy's structure and harbors most of the Galaxy's
star formation regions. The GRS sky coverage spans $18 \degr < l < 55
\degr$ and $\vert b\vert < 1 \degr$. Additional, incomplete sky coverage
is available for $14 \degr < l < 18 \degr$ over the same latitude range.
The GRS covers a total of 74 square degrees. The GRS maps the
distribution of emission from the $J = 1 \rightarrow 0$ ($\nu_0 = 110.2$
\ghz) rotational transition of \cor.  The \cor\ isotopologue is $\sim 50$ times
less abundant than \co\, and hence has a much smaller optical
depth. This decreased optical depth yields smaller line widths and
gives a cleaner separation of individual velocity components along any
specific line of sight compared to previous \co\ surveys.

The GRS, with a spectral resolution of $0.21 \kms$, an angular
resolution of $46\,\arcsec$, and a $22 \, \arcsec$ angular sampling,
improves upon all previous large scale CO surveys.  It is the only fully
sampled (in solid angle) large scale \cor\ survey extant. The GRS
improves upon the Bell Labs \cor\, survey \citep{lee01} that has a
spectral resolution of $0.68 \kms$, an angular resolution of $103
\arcsec$ and $180 \arcsec$ angular sampling.  The GRS also has better
resolution than previous \co\, surveys. For example, the University of
Massachusetts Stony Brook survey \citep{sanders86} has a spectral
resolution of $1.0 \kms$, an angular resolution of $45 \arcsec$ and
$180 \arcsec$ angular sampling. The Columbia/CfA \co\, survey
\citep{dame01} has a spectral resolution of $0.18 \kms$, an angular
resolution of 450\arcsec\, and $225\arcsec$ to $450 \arcsec$ angular
sampling. Compared to GRS, {\it all} of these surveys are severely
undersampled in angle. The GRS maps have spectra observed at positions
separated by $\sim \frac{1}{2}$ the telescope's beamwidth (0.48 HPBW,
actually). This, together with high spectral resolution allows us to
separate individual \cor\, components cleanly. Thus we can for the
first time study the molecular properties of Galactic \hii\ regions
free from angular sampling bias.

\section{H~{\bf\small II} Region Source Sample\label{sample}}

Here we study the molecular properties of all known \hii\ regions in the
zone mapped by the GRS. Our sample of 301 \hii\ regions contains 123 UC,
105 compact, and 73 diffuse nebulae.  For UC nebulae without RRL
measurements in the original papers, we compile RRL velocities from
\citet{afflerbach96} and \citet{araya02} or, if the UC and compact
sources were co-spatial, from L89.  The compact nebulae are from the L89
catalog. The majority of our diffuse regions are from L96, but a small
number are from the pilot survey of diffuse regions in L89.  This final
nebular sample results from further vetting of the \S 2 sources using
data from several recently completed Galactic scale sky surveys that
overlap the GRS zone.

We verify the existence, classification, and position of each \hii\
region by examining the radio continuum and infrared emission at its
nominal position. For the radio continuum emission, we primarily use
the 21\,cm VLA Galactic Plane survey
\citep[VGPS:][]{stil06}. In addition to the 21\,cm
\hi\ line emission data cubes, the VGPS generated 21cm continuum images
over the range $18\arcdeg < l < 67\arcdeg, |b| < 1-2$ with $\sim 1
\arcmin$ resolution.  In addition to the VLA measurements, the VGPS 
used a Green Bank Telescope 21\cm\ survey to provide the zero spacing data.  
The VGPS is therefore sensitive to both large and small scale emission.  We also use the 20\,cm data
from the Multi-Array Galactic Plane Imaging Survey
\citep[MAGPIS:][]{helfand06}. These data were collected with the VLA
operating in B-, C-, and D-configurations and have a resolution of
$\sim 6 \arcsec$.  The 20 \cm\ MAGPIS data cover the range $5\arcdeg <
l < 48\fdg 5, |b| < 0\fdg8$.  We find MAGPIS to be best suited for
verifying UC nebulae, while the VGPS is better suited for verifying
compact and diffuse \hii\ regions.  For the infrared emission, we use
the 8\,\micron\ data from the Galactic Legacy Infrared Mid-Plane
Survey Extraordinaire
\citep[GLIMPSE:][]{benjamin03} and 24\,\micron\ data from the MIPS Inner
Galactic Plane Survey (MIPSGAL: Carey et al. 2008 in preperation).
Both infrared surveys have coverage beyond the extent of the GRS.

For inclusion in our sample, we require a continuum peak at the
position of each \hii\ region. The UC regions were identified by their
IRAS colors and the compact and diffuse regions were located in radio
continuum maps with $2 \farcm 6$ resolution. Both of these
identification methods have some level of error that can be reduced
through correlation with high resolution radio continuum data. For UC
regions, continuum observations are necessary to confirm that the
nebula is an \hii\ region, and not a dense protostellar clump. The
sources from \citet{wc89a} and \citet{kurtz94} were confirmed to be UC
regions with VLA continuum observations, but the majority of sources
from \citet{watson03} and \citet{sewilo04} have not yet been confirmed
with high resolution ratio data. We exclude 11 UC sources, 2 compact
sources and 2 diffuse sources that do not have significant continuum
emission.

We remove all known SNR from our sample by comparing our positions
with the catalog of \citet{green06}\footnote[2]{Available at
http://www.mrao.cam.ac.uk/surveys/snrs/)} as well as with two recent
catalogs of SNRs by \citet{brogan06} and \citet{helfand06}. Both of
these recent catalogs compute the spectral index of SNR candidates using
20\,\cm\ and 90\,\cm\ VLA continuum data and rely on the anticorrelation
of SNRs with infrared emission (8\,\micron\ {\it MSX} data in the case of
\citet{brogan06} and 21\,\micron\ {\it MSX} data for \citet{helfand06}). We
exclude 13 SNRs from our sample that were found in these catalogs, as
well as one found in \citet{gaensler99}. There are a comparable number
of sources that are spatially coincident with SNRs, but which have a
strong IR component. We believe these are \hii\ regions in locations
that have produced multiple generations of stars.  These sources are
retained in our sample.  We also remove an additional six sources that
do not have infrared emission since they most likely are non-thermal.

Finally, we determine if the classification (UC, compact, diffuse) and
position of each source are correct, and remove duplicate sources. We
require UC \hii\ regions to have small ($\lesssim 1\arcmin$) bright knots of continuum
emission, compact regions to be larger bright continuum sources, and
diffuse regions to have faint extended continuum emission. There are
many cases where an UC region was mistakenly identified as a compact or
diffuse region in the L89 and L96 catalogs. This misidentification is
due to the fact that UCs are unresolved with the 2\farcm6 beam of
\citet{altenhoff79} from which L89 and L96 drew their positions. We
exclude 69 compact and diffuse nebulae whose positions are coincident
with UC regions.  Many of the UC \hii\ regions found in the GRS are
in large complexes with individual UC components separated by angular
distances less than the $46\arcsec$ GRS beam. We treat these
complexes as single UC regions because they share a common molecular gas
clump at the GRS resolution.

\begin{figure}
\epsscale{1.0}
\plotone{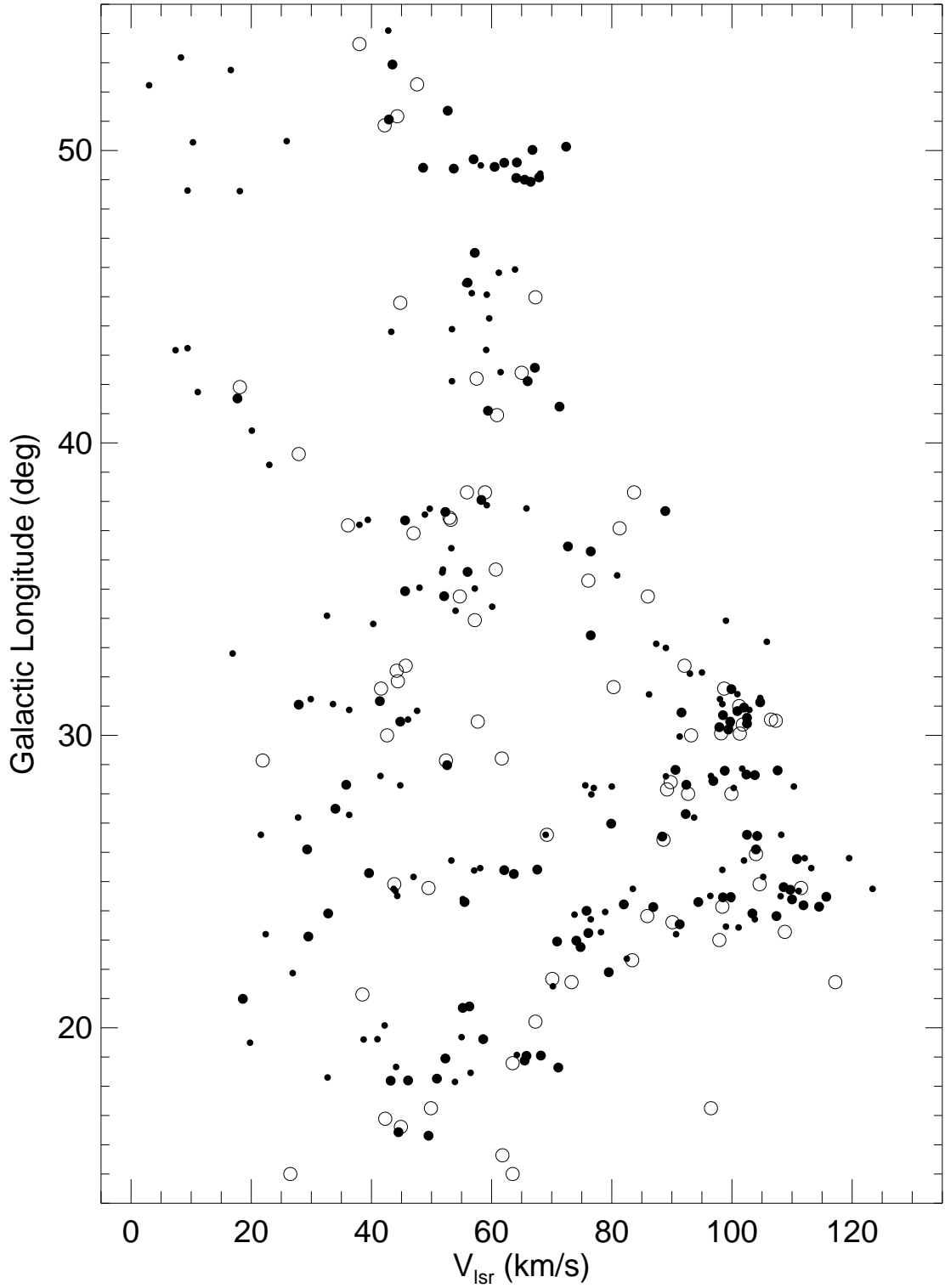}

\caption{Longitude-LSR velocity diagram for \hii\ regions located inside
the GRS survey zone. The nebulae are shown projected onto the Galactic
plane.  The symbols represent UC nebulae (small filled circles), compact
nebulae (medium filled circles), and diffuse nebulae (large open circles).}

\label{fig:lv}
\end{figure}

Table \ref{tab:anomalous} lists the 38 \hii\ regions we cull from our
sample because of the criteria just described.  Table
\ref{tab:anomalous} gives the source name, the reason for exclusion from
our sample, and the reference if the source is a known SNR. Source names
identify the nebular type: UC (``U''), compact (``C''), or diffuse
(``D'') . This source name convention will be followed throughout this
paper.
Figure \ref{fig:lv} shows the longitude-velocity position of our
nebular sample. The symbols in Figure \ref{fig:lv} indicate the
nebular type: UC (small filled circles), compact (medium filled
circles), or diffuse (large open circles).  Table \ref{tab:diffuse}
gives the properties of the nebulae in our sample.  Listed are the
source name, its position in Galactic and equatorial coordinates and
its RRL velocity\footnote[3]{The RRL velocities here are in the
kinematic Local Standard of Rest (LSR) frame using the radio
definition of the Doppler shift.  The kinematic LSR is defined by a
solar motion of 20.0 \kms\ toward ($\alpha, \delta$) = (18$^{\rm h},
+30\arcdeg$)[1900.0].} with its $1 \sigma$ error.  Altogether our 301
\hii\ region sources probe 266 unique directions since 33 nebulae have
RRL emission at several different velocities that presumably
originates from physically distinct nebulae located along the line of
sight.  For some \hii\ regions, we change the classification based on
the morphology of the VGPS and MAGPIS radio continuum emission.  We
also change the position of a few \hii\ regions based on this
continuum emission if the position is obviously incorrect.  All such
changes are noted with footnotes in Table \ref{tab:diffuse}.

\section{Finding Molecular Gas Associated with Galactic H~{\bf\small II} Regions\label{sec:ID}}

To characterize the properties of molecular gas associated with Galactic
\hii\ regions we must establish reliable morphological correlations in
({\it l, b, V\/})--space between the \cor\ gas and the nebulae. The
GRS is a large ($\sim10 ~{\rm Gbyte}$), dataset that contains very
complex \cor\ emission line structure.  The ($l, b, V_{\rm LSR}$)
structure of the molecular gas near the nebulae can be very
complicated.  Any given line of sight frequently contains multiple
emission lines.  There probably is no single algorithm that can
uniquely give reliable CO/\hii\ regoim ({\it l, b, V\/}) morphological
correlations. We therefore use a suite of software tools to analyze
the GRS data cubes in a variety of ways.

\subsection{Analysis Software \label{sec:software}}

To maximize the power and flexibility of our analysis we wrote a large
suite of IDL procedures for spectral analysis rather than use any of the
standard single-dish radio astronomy software packages.  We used the
single-dish radio astronomy TMBIDL\footnote[4]{Written by T. M. Bania
and available at http://www.bu.edu/iar/research/dapsdr/} software as our
starting point.  TMBIDL was originally written to analyze NRAO Green
Bank Telescope data. (TMBIDL was the inspiration that led to the NRAO
GBTIDL\footnote[5]{See http://gbtidl.sourceforge.net/} software.)  The
TMBIDL software emulates and improves upon many of the features of the
NRAO UniPops analysis program.  It includes Gaussian and polynomial line
fitting, data visualization, data manipulation, etc.  The TMBIDL code
can easily be modified to analyze data from any single dish radio
telescope.

We wrote additional IDL procedures to interface and analyze GRS data
within the TMBIDL environment. TMBIDL was created to analyze single
spectra, so we added ({\it l-V\/}), ({\it b-V\/}), and ({\it l-b}) mapping
tools to better visualize the GRS ($l, b, V$) data cubes.  The basic
visualization is a normalized contour map of the CO emission.  Using
these tools, we found that the morphology of the CO emission at the
positions of the \hii\ regions was often very complex.

To gain further control over the visualization of the GRS data, 
we wrote GUI-based software\footnote[6]{Available for download at
http://people.bu.edu/andersld/}
to analyze images extracted from the GRS ({\it l,b,V\/}) FITS data
cubes.  This software provides powerful GUI tools to extract, image, and
analyze subcubes for each sample \hii\ region.  For our analysis we
imaged various quantities for an ({\it l,b}) zone surrounding each
nebula.  The user can, for example, easily modify the velocity channel
whose \cor\ line intensity is being imaged over the mapped region.  One
can also quickly create and display integrated intensity CO maps, \wco\
[K \kms], as well as arbitrarily vary the velocity range of the
integration, $\Delta V$.  (This is done with a kernel based algorithm
that is fast and efficient.)  Sub-images and regions can be created for
any image, saved and then reloaded at any time. The GUI uses 
DS9\footnote[7]{Available for download at
http://hea-www.harvard.edu/RD/ds9/} syntax to define regions.   
These regions can have all the basic DS9 shapes.  We added a
``threshold'' region that selects all contiguous pixels above a
user-defined threshold level that surround a given image pixel.  Using
this thresholding tool one can identify and analyze arbitrarily
complex morphologies.

This GUI-software is fully integrated with TMBIDL.  The user can, for
example, export the spectra within any region to TMBIDL for spectral
analysis.  The ability to scan quickly through velocity channels, create
integrated intensity images, select pixels and fit Gaussians to spectra
--- all within a single application --- is extremely powerful.

\subsection{Correlation of Molecular Gas and H~{\bf\small II} Regions}

Our goal is to find a morphological coincidence in ({\it l,b,V\/}) --
space between the CO gas and the \hii\ region. After a coincidence is
established, we want to characterize the molecular gas using spectral
fits to the \cor\ emission. The UC positions are in general known to
an accuracy greater than the $22 \arcsec$ GRS pixel spacing. For the
compact and diffuse \hii\ regions, our positions are accurate to a few
arcminutes.  The RRL LSR velocities are accurate to $\sim\, 0.1 \kms$
(from Gaussian fits).  Although the ({\it l,b,V\/}) position of each nebula is
accurately known, establishing a robust set of criteria for
identifying a real molecular/\hii\ physical association is nontrivial.

The CO emission maps of \hii\ regions can be quite complex due to GMC
structure and PDR/ionization front interactions.  For example, one
expects the molecular and ionized gas velocities to diverge as an \hii\
region evolves.  Once an OB star forms within a GMC its ionization front
(IF) expands rapidly at first, reaching the nebular Str\"{o}mgren radius
in $\lesssim 10^5$ yr.  The IF pushes the surrounding GMC molecular gas
outwards.  \citet{dyson97} show that as the IF expands, it rapidly slows
until it is expanding at $\sim 10 \kms$ when it reaches the
Str\"{o}mgren radius.  Since \hii\ regions are generally older than
$10^5$ yr (with the possible exception of some UC \hii\ regions), the
maximum difference between the RRL and the associated molecular gas
should be $\lesssim 10 \kms$.

Our nebular sample has \hii\ regions of different ages and thus should
show evolutionary effects.  Diffuse \hii\ regions should be older than the UC
nebulae, and thus have had more time to evolve away from and displace their natal
clouds. The molecular gas in diffuse \hii\ regions should show a weaker
association or may not be present at all. We expect the majority of UC
and compact \hii\ regions to be associated with a molecular clump
\citep[see, e.g.,][]{kim03}. Because of these complications, our
CO/\hii\ analysis is comprised of a series of distinct investigations.

\subsubsection{Single Position Spectrum Analysis}\label{sec:singlepointing}

We first examine the GRS \cor\ spectrum at the nominal position of each
\hii\ region. Most previous studies of the molecular component of \hii\ regions
were made using single pointings, e.g., \citet{whiteoak82}; Russiel \&
Castets (2004).  All report that the majority of \hii\ regions have
associated CO.  \citet{whiteoak82} found in their survey of \co\
emission from Southern \hii\ regions that molecular gas within 5 \kms\
of the RRL velocity had large line intensities, and therefore was
probably associated with the \hii\ region. In their analysis of Southern
compact \hii\ regions using both \co\ and \cor, \citet{russeil04} argued
that 10 \kms\ is a better criterion for determining a molecular/\hii\
association.

To make a single pointing CO/\hii\ comparison we calculate an average
\cor\ spectrum at the position of each \hii\ region by convolving the
GRS datacube (which is oversampled in angle) with the FCRAO telescope
beam (HPBW = $46 \arcsec$).  We then search this average spectrum for
a \cor\ emission line peak at the \hii\ region RRL velocity.
Specifically, we look for emission above 0.5\K\ brightness
temperature, $T_{\rm MB}$, and within $\pm 2.5 \kms$ of the \hii\
region LSR velocity.

Our brightness temperature limit is chosen to be well above the GRS
noise.  The GRS data have a typical RMS sensitivity of $\sigma({\rm
T_{MB}})=0.27\K$.  
The beam convolved average spectrum has a factor of $\sim\,2$ decrease
in noise compared to a single GRS position.  Thus our 0.5\K\
search criterion is a $\sim3\sigma$ limit.

Using the 0.5\,K and $\pm 2.5\kms$ criteria, only 52 \% of the nebular
sample shows a CO/\hii\ association.  Repeating this procedure with the
same intensity requirement, but with velocity ranges of $\pm 5.0 \kms$
and $\pm 7.5 \kms$, we find that, respectively, $70\%$ and $79\%$ of
\hii\ regions meet these criteria.  Certainly increasing the velocity
range further still will yield a greater number of nebulae matching the
association criteria, but relaxing the association definition in this
way also increases the chance of a misidentification and the possibility
of blending multiple velocity components.

\subsubsection{\wco\, Integrated Intensity Map Analysis}\label{sec:map_analysis}

We use the GRS data to make an ({\it l, b\/}) contour map of the \cor\
integrated intensity, \wco\ (K \kms), in order to provide information
about the spatial distribution of the molecular gas.  The main
weakness of the single pointing method of searching for CO/\hii\
region associations is that sources with molecular gas offset from the
nominal position of the \hii\ region are not counted as detections.
For each nebula we use TMBIDL to make normalized \wco\ contour maps
that are ${\it l} \times {\it b} = 12 \arcmin \times 12 \arcmin$
($\sim\,30 \times 30$ GRS pixels) in size.  The map \wco\ is
calculated by integrating each spectrum over $15 \kms$ centered at the RRL velocity.  The map peak \wco\
is used to normalize the \wco\ value for each pixel.  The final
product is a normalized contour map for each nebula in the sample.

We then search each map for \wco\, peaks and note the distance from
the 80\% peak \wco\ contour to the nominal \hii\ region position.  We
define any source where this distance lies within the \hii\ region
positional error bars to be a positive detection and any source where
this distance is just outside of the error bars (roughly twice the
positional uncertainty) to be an ambiguous detection.  All sources not
meeting these positional criteria are deemed to be non-detections. We
find that 70\% of our sample nebulae show positive detections, 14\%
have ambiguous detections, and 16\% show no correlation between the
\hii\ region position and the \cor\ emission.  Somewhat suprisingly,
adding the molecular spatial distribution information to the CO/\hii\
association criterion did not add significantly to the detection rate.

\subsubsection{CO ($\/l,b,V$) Data Cube Analysis Procedure}\label{sec:gui}

\begin{figure}
\epsscale{1.0} 
\plotone{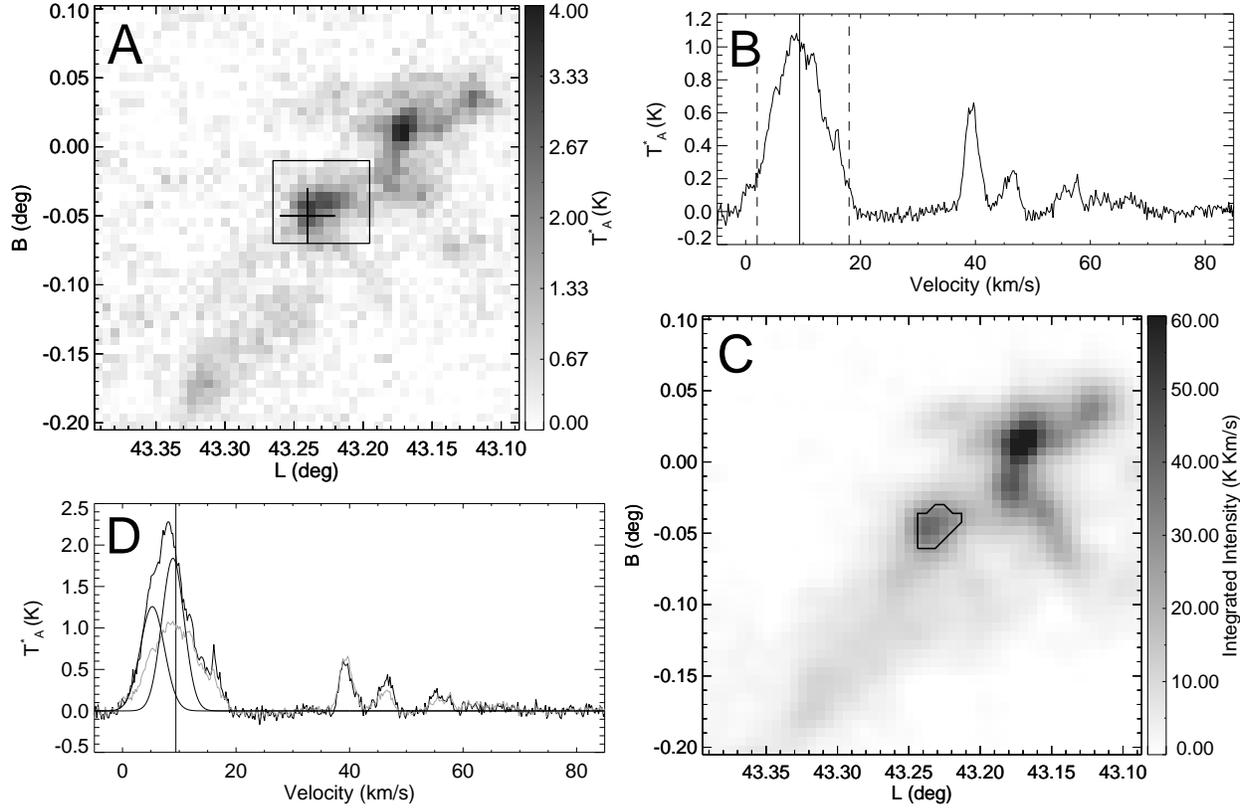}
\caption{The CO/\hii\ association procedure for the U43.24$-$0.05 
\hii\ region (see text). 
Panel A: Using single velocity channels near the RRL velocity and
nominal \hii\ region position we first locate the associated molecular
emission.  The black cross shows the \hii\ region position.  The black
rectangle defines the extent of the \cor\ emission that is deemed to be
associated with the \hii\ region.
Panel B: The average \cor\ spectrum of the GRS data cube voxels that lie
within the panel A black rectange is shown.  The solid vertical line flags the
\hii\ region RRL velocity and the dashed lines show the velocity range of the associated emission.
Panel C: The integrated intensity image made from the data cube using
the line center velocity and FWHM line width found by the Panel B
Gaussian fit is shown.  This image is then used to define the final
extent of \cor\ emission that is deemed to be associated with the \hii\
region.  The black outline shows this region which is defined by a
threshold algorithm (see text).
Panel D: The average \cor\ spectrum of the data cube voxels that lie
within the panel C threshold defined region is shown. The vertical line
flags the RRL velocity.  The  Gaussian fit to the emission
line is used to derive the physical properties of the \cor\ gas
associated with this \hii\ region.  The spectrum from panel B is shown in gray for comparison.
}

\label{fig:3plot}
\end{figure}

Clearly, the molecular emission in the GRS is complicated and
difficult to characterize.  These experiments in establishing a
CO/\hii\ region association demonstrate the need for a more
sophisticated analysis. We use the \S \ref{sec:software} GUI software
to search the GRS \cor\, data cubes in ($l, b, V$) parameter space.
For each nebula we make a series of ($l,b$) \wco\ images and search
for CO/\hii\ region associations.  We follow the four step iterative
procedure described below and illustrated in Figure \ref{fig:3plot}.

\begin{enumerate}
\item  We first find the velocity range of the molecular emission 
associated with the \hii\ region. We examine single velocity channel
images (position-position ({\it l, b\/}) maps) centered at the nominal
position of the \hii\ region at its RRL velocity with overlaid VGPS
21cm continuum contours.  We scan through these single channel images
over $\pm 10 \kms$ of the source RRL velocity, searching for the
channel where the molecular emission at the position of the \hii\
region has the highest intensity. If we are able to identify a
molecular clump near the position of the \hii\ region, we extract the
spectra from the voxels near this molecular clump.  We fit a Gaussian
to the (unweighted) average spectrum of this extracted emission and
record the line center and FWHM of the emission line.  Frequently, the
molecular emission at the \hii\ region position is either absent or
has a morphology that is difficult to characterize from single channel
maps. For $\sim\,25\%$ of our nebulae we are unable to make a
molecular gas association from these single channel maps.

\item Next we make an integrated intensity map, \wco, by summing the
intensities at a given ({\it l,b}) over the range of velocities found in
step (1). If the source has an unambiguous CO/\hii\ association, we
calculate \wco\, over the velocity range $V \pm \Delta V/2$, where $V$
is the step (1) Gaussian line velocity and $\Delta V$ is its FWHM line
width.  If the source has an ambiguous step (1) association, we
calculate \wco\, centered at the source RRL velocity over the range $\pm
10 \kms$.

\item We then use the $W_{\rm CO}$ image created in step (2) to
find pixels with molecular emission associated with the \hii\ region.
We find the brightest emission near the \hii\ region in the \wco\,
image and select all contiguous pixels that have values above a
threshold determined independently for each \hii\ region.  The
threshold is varied until a small number of pixels are selected ---
typically 20 to 30.  The exact number is set by the molecular clump's
intensity profile.  Small clumps with a sharply peaked intensity
profile have fewer pixels, whereas larger clumps with a ``plateau'' of
emission have more pixels.  This method seeks to isolate single
clumps in order to preserve the line peak intensity and minimize
blending of multiple velocity components.

\item All the GRS spectra within the ($l,b$) region selected in step (3)
are used to calculate an average \cor\, spectrum.  We then fit the
fewest possible number of Gaussian components to the spectrum in order
to maximize the intensity of the peak CO emission.  The majority
($\sim\,80\%$) of our sources are adequately fit with only one
Gaussian component. Many nebulae, however, do not have clean
Gaussians profiles, but rather show structure that often suggests a fainter,
wider line superposed on a brighter, narrower line.  
\end{enumerate}

Our analysis procedure is summarized in Figure \ref{fig:3plot} for the
\hii\ region U43.24$-$0.05. Panels A and B depict step 1. Panel A
shows a single channel GRS map at the velocity nearest the RRL
velocity where the molecular clump has the highest intensity.  The
cross marks the nominal \hii\ region position.  For clarity we have
not shown the VGPS continuum emission contours.  The black box shows
the ($l,b$) positions of the voxels from which we extract spectra to
produce the average spectrum shown in Panel B. The vertical line in
Panels B and D marks the RRL velocity.  Using the velocity range of
the emission line shown as dashed lines in Panel B, we create the
integrated intensity image shown in Panel C (step 2). This image has
been smoothed with a $3 \times 3$ Gaussian kernel.  The black outline in Panel C shows contiguous
integrated intensity values above a threshold (step 3).  This region
and its threshhold is defined by visual inspection of the image.  This
is the \cor\ emission we deem to be associated with the \hii\
region.  Finally, we extract the spectra from these ($l,b$) positions
to produce the average spectrum shown in Panel D (step 4).  For this
source, our method separates the two emission lines that are blended
in the panel B spectrum and preserves the peak line intensity.


In principle, we could iterate our analysis to locate and fit the
molecular emission more accurately.  Using the Panel D Gaussian fit, we
could create a new integrated intensity image, define a new CO/\hii\
association region, and fit a Gaussian to this new average spectrum.  We
did this for 10 test cases and found only only minimal changes that were
not significantly different from a single pass analysis. 

This procedure has many advantages.  We are able to characterize the
spectral properties of molecular gas distributions that have arbitrary
morphologies.  We minimize our assumptions at every stage of the
process.  By first examining the ({\it l,b}) images of each source at
individual velocity channels, we limit false detections that may arise
from integrated intensity images containing multiple velocity
components blended together.  By extracting the spectra from regions
defined in the integrated intensity images, we make no assumptions
based on the visual appearance of the molecular emission at individual
velocity channels.  By using a variable threshold to select pixels, we
are able to characterize molecular structures with arbitrary
morphology.  This threshold definition ensures clean spectral fitting;
the spectra are not contaminated by adjacent pixels that would lower
the line intensity and might increase the width of the fitted Gaussian
line.  Finally, since we analyze an average of many spectra, the lines
we fit have a much greater signal to noise ratio than a single
pointing spectrum.  The Gaussian fit uncertainties in the spectral
line parameters we derive are thus minimized.

After we establish a CO/\hii\ region association, we characterize the
angular size of the molecular cloud by fitting an ellipse to the \wco\
spatial distribution.  Since the association defined in Panel C
of Figure \ref{fig:3plot} is a \wco\ threshold value that is
determined independently for each source, the size of an ellipse
fitted to this zone would have little physical meaning.  We therefore
define a new threshold selected region that is uniform for
the entire nebular sample.  This new region is defined by a threshold
set to 80\% of the \wco\ peak inside the original association zone.
The fitting algorithm calculates the semimajor and semiminor axes
using the ``mass density'' of pixel locations: the clustering of pixel
locations along the {\it l} and {\it b} directions.  The fitted
ellipse to this region is thus a uniform estimate of the size of the
molecular gas associated with the \hii\ region.

The properties we derive for the CO/\hii\ region associations are
summarized in Table \ref{tab:properties}.  For each nebula we list the
parameters of the ellipse and Gaussian line fits. All errors are $1
\sigma$ uncertainties.  Given are the ellipse centroid in Galactic
coordinates, its size, semi-major, $a$, and semi-minor, $b$, axes,
together with the position angle measured from north toward increasing
Galactic longitude.  The ellipse size is the geometric mean diameter,
$2\sqrt{a b}$.  If there are multiple spectral components in the
Gaussian fit only the properties of the brightest are listed.  We also
use only this brightest component in our subsequent analysis.  The
fitted line parameters given are the center velocity, intensity (in
main beam brightness temperature units), and FWHM line width, $\Delta
V$. Table \ref{tab:properties} also lists some additional nebular
properties derived below in \S \ref{sec:discussion}: the CO excitation
temperature, the \cor\ column density, and the nebular confidence
parameter, $CP$.

\section{Discussion \label{sec:discussion}}

Our analysis here provides a large sample of molecular cloud/\hii\
region associations whose physical properties are well characterized.
In \S\ref{sec:ID} we show that the molecular emission is often
morphologically complex and offset from the nominal \hii\ region
position and RRL velocity.  We find that the traditional single
pointing analysis does not reliably detect the molecular components of
\hii\ regions.  It is very difficult to make a CO/\hii\ association with any
confidence without an image produced from a $(l, b, V)$ data cube that
spans the entire \hii\ region/PDR/molecular cloud interaction region.


To distinguish sources with an unambiguous molecular gas component at
the \hii\ region position and velocity from those with less robust
molecular gas associations, we assign a confidence parameter, $CP$, to
each source ranging from A to E.  Our qualitative criteria for the
confidence parameter are as follows: 
A: no ambiguity in position or velocity, the molecular gas coincident 
with radio continuum or shows clear signs of interaction; 
B: either offset somewhat in position or velocity, or in a
complex region of molecular gas; 
C: either offset in position or velocity or in a complex region, 
but fainter than a B source and offset further; 
D: diffuse emission near the correct position and velocity, but 
uncharacterizable due to low intensity or ambiguous morphology; 
E: nothing at all apparent in position and velocity. 

\begin{figure}
\epsscale{1.0} 
\plotone{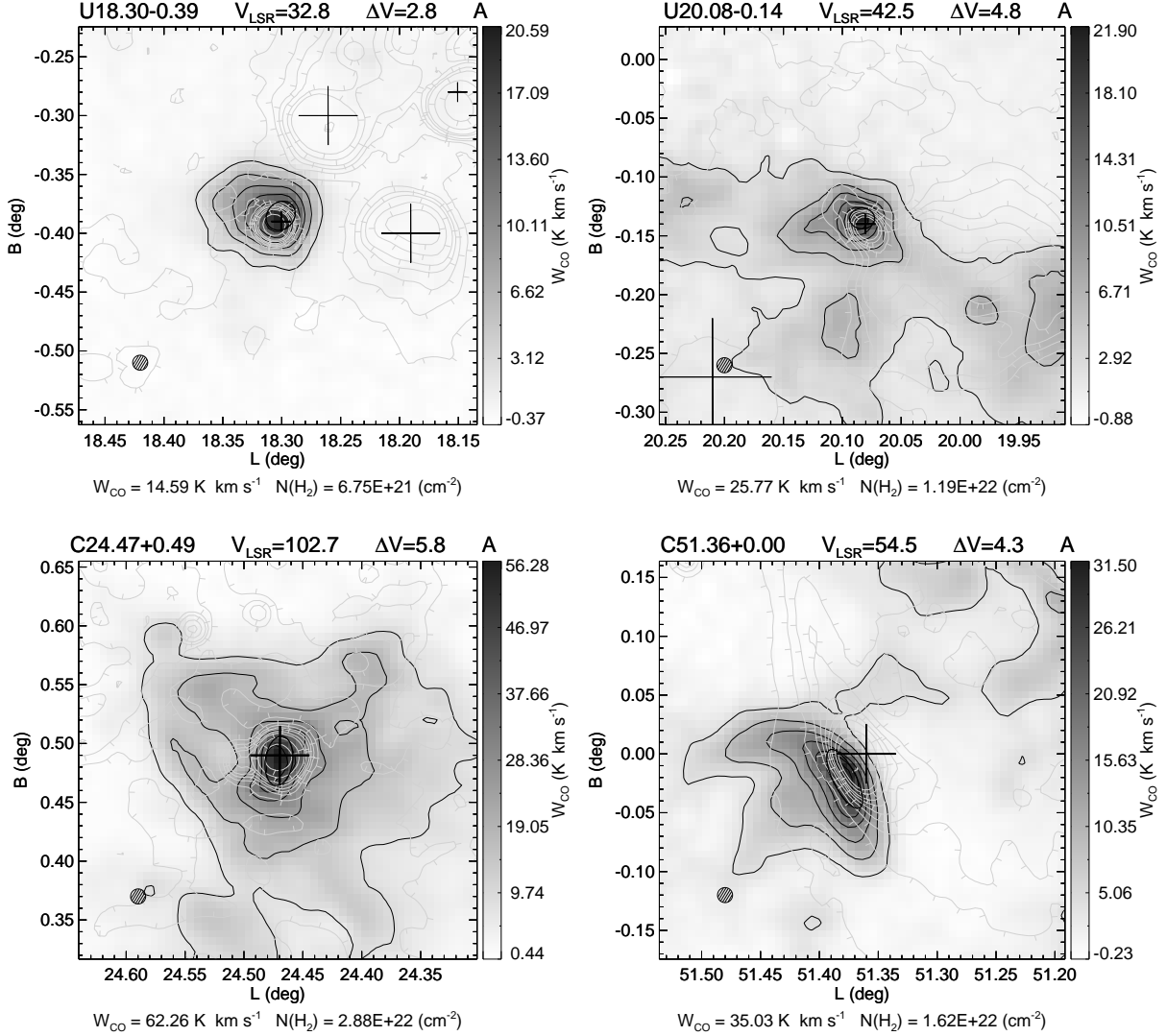} 

\caption{Integrated intensity images (grey scale and black contours) 
for a sample of \hii\ regions with A confidence parameters. Nominal
\hii\ region positions are marked with crosses and our fitted ellipses
are also shown.  Contours of 21cm VGPS continuum emission are
shown in grey.  Tickmarks on the VGPS contours point downhill, towards decreasing values.  See
\S \ref{sec:discussion} for a detailed description of these images.}

\label{fig:images_a}
\end{figure}

\begin{figure}
\epsscale{1.0} 
\plotone{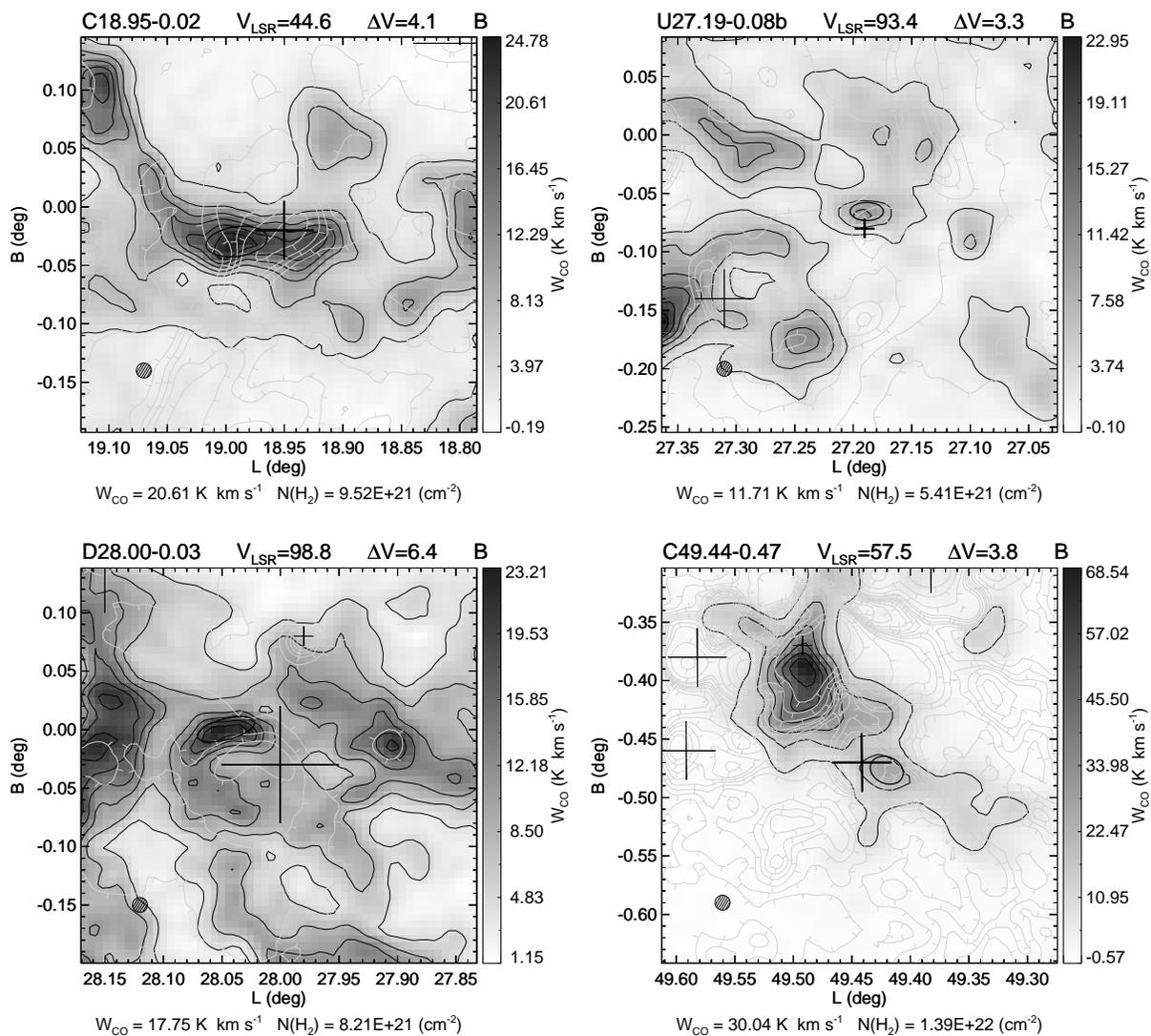}  

\caption{Integrated intensity images for a sample of \hii\ regions with
B confidence parameters.  Other parameters are as in Figure \ref{fig:images_a}.}

\label{fig:images_b}
\end{figure}

\begin{figure}
\epsscale{1.0} 
\plotone{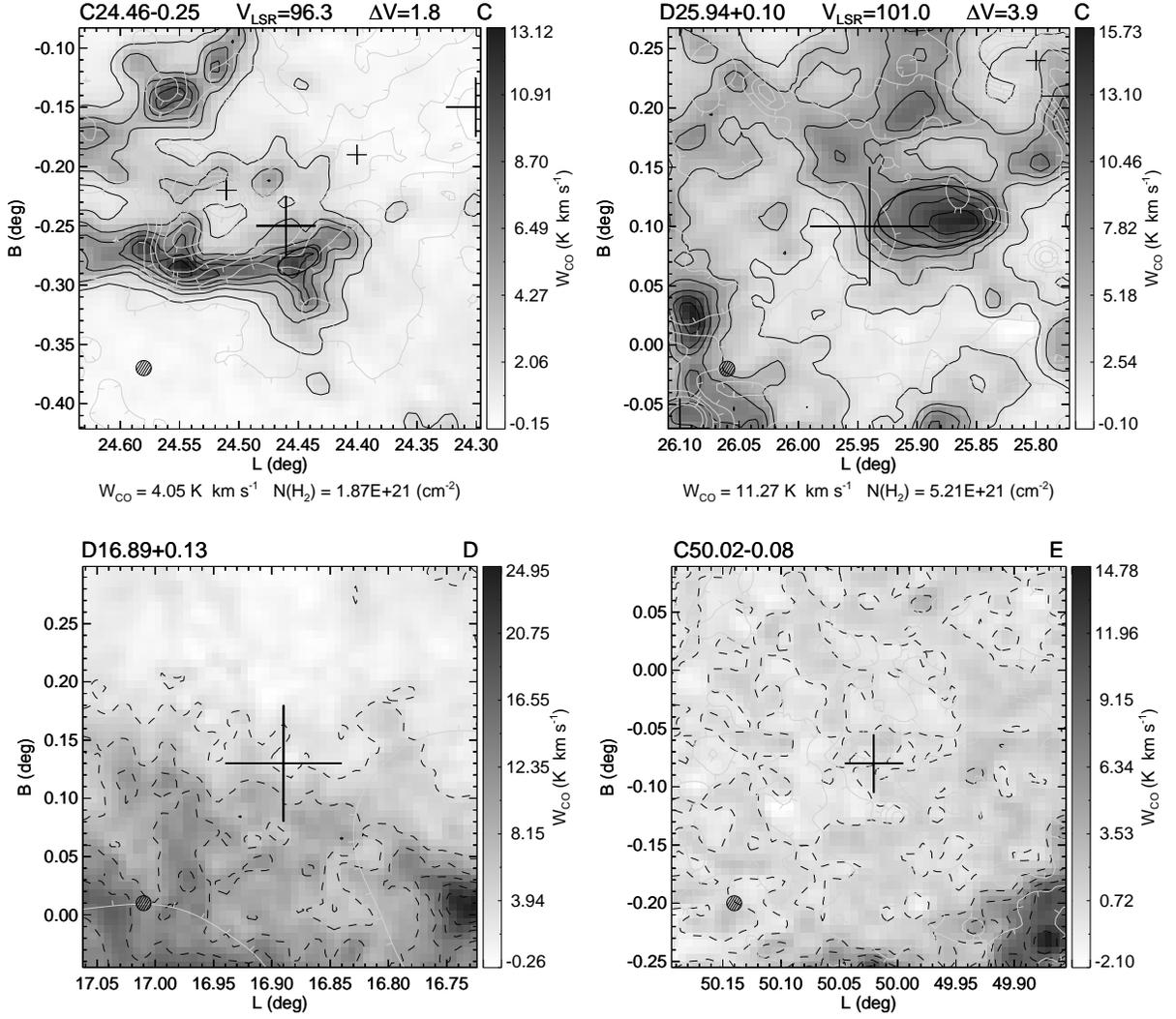}  

\caption{Integrated intensity images for a sample of \hii\ regions with
C confidence parameters (top row), D (lower left) and E (lower
right). Fitted ellipses are shown for C sources only.  Other
parameters are as in Figure \ref{fig:images_a}.}

\label{fig:images_cde}
\end{figure}

Figures \ref{fig:images_a}, \ref{fig:images_b} and
\ref{fig:images_cde} give representative examples of our confidence
parameter classification.  These $20 \arcmin \times 20 \arcmin$ \cor\,
integrated intensity images span the range of molecular gas
morphologies surrounding our sample of \hii\ regions.  The images are
grouped by their confidence parameter classification: Figure
\ref{fig:images_a} shows $CP$ A sources, Figure \ref{fig:images_b}
shows $CP$ B sources and Figure \ref{fig:images_cde} shows $CP$ C, D,
and E sources.  Because the GRS is oversampled ($22\arcsec$ pixels at
$46\arcsec$ HPBW), we increase the signal to noise by smoothing the
images with a $3 \times 3$ Gaussian filter.

At the top of each image we list the \hii\ region name, the \cor\ line
velocity (\kms) and FWHM line width (\kms), $\Delta V$, and the source's confidence
parameter.  These Gaussian fitted line parameters are used to
calculate the source's \cor\ integrated intensity, \wco (K \kms), for
the velocity range $V \pm \Delta V /2$.  At the bottom of each image
we give the \wco\, calculated from the fitted line: $W_{\rm CO} = 1.06
\, T_{\rm MB} \, \Delta v$, which is accurate for Gaussian line
shapes.  Using the procedure described in \citet{simon01}, we also
give an estimate of the ${\rm H_2}$ column density, $N({\rm H_2})
[{\rm cm^{-2}}]$.

The grey scale image shows the \wco\ distribution.  Normalized
contours are drawn at 83\%, 67\%, 50\%, 33\% and 16\% of this maximum.
(One can get quantitive \wco\ values using the scale bar at the
right.)  For \hii\ regions where we could not associate molecular gas,
the contour lines are dashed rather than solid and, of course, no line
parameters are given.

A bold cross marks the nominal position of the \hii\ region.  Plotted
in bold is the fitted ellipse described in \S \ref{sec:gui}.  If there
are other \hii\ regions in the field, they are marked with thinner
crosses. The cross arm lengths correspond to the beam size used to
make the measurement of the \hii\ region ($3 \arcmin$ for L89;
$9\arcmin$ for L96), except for the UC regions where the cross arm
lengths are set to 1\arcmin.  The GRS beam (HPBW = $46
\arcsec$) is shown in the lower left corner of each
image. Shown in grey in these images are VGPS 21cm continuum contours.
Tickmarks on these contours point downhill, towards decreasing 21cm
emission.

\begin{figure}
\epsscale{1.0} 
\plotone{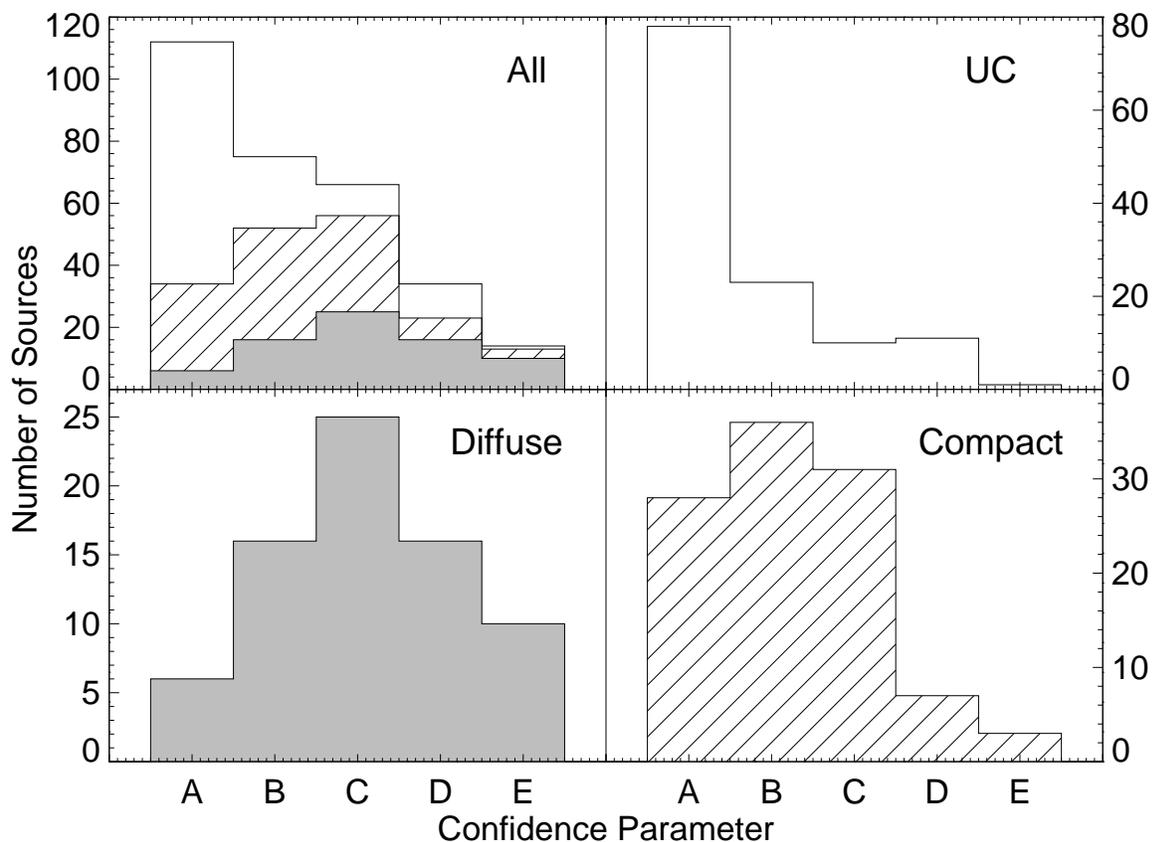}

\caption{The number distributions of the qualitative confidence
parameter, $CP$.  Sources with associated molecular gas are rated A, B
or C in order of decreasing confidence in the association (see \S \ref{sec:discussion} text). Confidence
parameter D and E sources have little or no associated molecular
emission, respectively. The top left panel is a stacked histogram
where the top line shows the histogram for the entire sample of 301
\hii\ regions.  The open, hatched and grey histograms show the
contribution that UC, Compact, and Diffuse nebulae, respectively, make
to the total in each bin.}

\label{fig:confidence}
\end{figure}

We are able to establish a highly confident ($CP$ A and B) CO/\hii\
region association for 62\% of the nebulae in our sample.  Relaxing
the confidence criterion to $CP$ A, B and C sources gives CO/\hii\
associations for 84\% of our nebular sample.  Histograms of the number
distribution of confidence parameter values are shown in Figure
\ref{fig:confidence} for UC (open), compact (hatched), and diffuse
(gray) nebulae.  The top left panel is the stacked histogram of the
distribution; the top line represents the entire sample.  For example,
there are a total of 113 $CP$ A sources: 78 UC, 29 compact, and 6
diffuse \hii\ regions.  Clockwise from here are the individual
histograms for the UC, compact, and diffuse nebulae.  (Most subsequent
figures will follow this display format.)  As is clear from Figure
\ref{fig:confidence}, the UC sample has the greatest number of high
confidence CO/\hii\ region associations.  

Some \hii\ regions appear to have no molecular gas associated with
them. Of our 301 sources, 14 (5\%) are classified as E sources; these
nebulae show no \cor\, emission whatsoever.  Thirty four sources
(11\%) are classified as D and therefore have only diffuse emission at
the correct position and velocity.  Five of our E sources and 15 of
our D sources have multiple RRL velocities along the same line of
sight. In these cases, one RRL velocity is probably from the \hii\
region of interest while the other is likely from a nearby \hii\
region.  Eleven of the 12 UC regions with confidence parameter values
of D or E have multiple velocity components.  

Two UC \hii\ regions are worth mentioning individually because of the
nature of their molecular associations.  The UC \hii\ region
U33.13+0.09 has a very large velocity offset between the molecular
material and the RRL velocity.  L89 find a RRL velocity of 93.8 \kms\,
and \citet{araya02} find a RRL velocity of 87.4 \kms.  Our \cor\
velocity of 75 \kms\ is in agreement with the CS velocity found by
\citet{bronfman96}.  The morphology and linewidth of the molecular
emission suggest that the molecular gas is associated with the \hii\
region.  The UC \hii\ region U21.42$-$0.54 also has a compact
molecular clump at the correct ({\it l, b}\/) position.  The \cor\
velocity of this clump, 54 \kms, is 16 \kms\ offset from its RRL
velocity of 70 \kms.  This source was not detected by
\citet{bronfman96}.  We assign a confidence parameter value of 
C to U21.42$-$0.54 because of the extreme velocity offset.

The lack of CO/\hii\ region associations in $\sim\,20\%$ of our sample
is not entirely unexpected.  This has been reported in the literature
before, although previous studies did not have datasets that were
fully sampled in angle as is the GRS.  \citet{blitz82} found a lack of
associations in $30\%$ of the Sharpless \hii\ regions studied.
\citet{russeil04} found no association with CO in $\sim 20
\%$ of their sample of Southern compact \hii\ regions.
\citet{churchwell90} do not detect the dense gas tracer ammonia in
$\sim\,30\%$ of a sample of 84 UC \hii\ regions.

Our large sample of \hii\ regions contains nebulae spanning a range of
evolutionary stages. The UC regions, being young, are more likely to
lie within their natal molecular clouds where the density of \cor\,
should be high.  As the \hii\ region evolves, it will dissipate the
gas.  We should see evidence for this in the compact and diffuse
regions.  We therefore expect UC regions to be associated with small
sizes, high excitation temperatures and column densities, large line
widths, and bright line intensities. Diffuse regions should have large
sizes and smaller values for the other quantities compared to the UC
regions. Compact regions should lie in between these two extremes.

The lack of molecular gas in an \hii\ region can certainly be an
evolutionary effect.  These nebulae may represent an older population
of \hii\ regions that have had time to become displaced from the gas
in which they formed due to a variety of mechanisms including, stellar
winds, ionization fronts, high stellar space velocities, etc.  One
observational consequence of this scenario might be a bubble
morphology in the molecular gas, seen as a ring in projection.
\citet{churchwell06} found a large population of bubbles in the
GLIMPSE survey \citep{benjamin03}.  Observed at mid-infrared
wavelengths, GLIMPSE boasts an angular resolution 10 times that of the
GRS and is therefore a better diagnostic tool for locating bubble
features.  We indeed see bubbles in \cor\, in most \hii\ regions in
the GRS for sources where \citet{churchwell06} also identified
bubbles. These regions are usually classified as D or E sources
because the gas has been pushed far away from the center of the
continuum emission. Further investigation of this topic will be the
subject of a future paper.

\subsection{Properties of Molecular Cloud/H~{\bf\small II} Region Sources 
\label{sec:properties}}

\begin{figure}
\epsscale{0.90}
\plotone{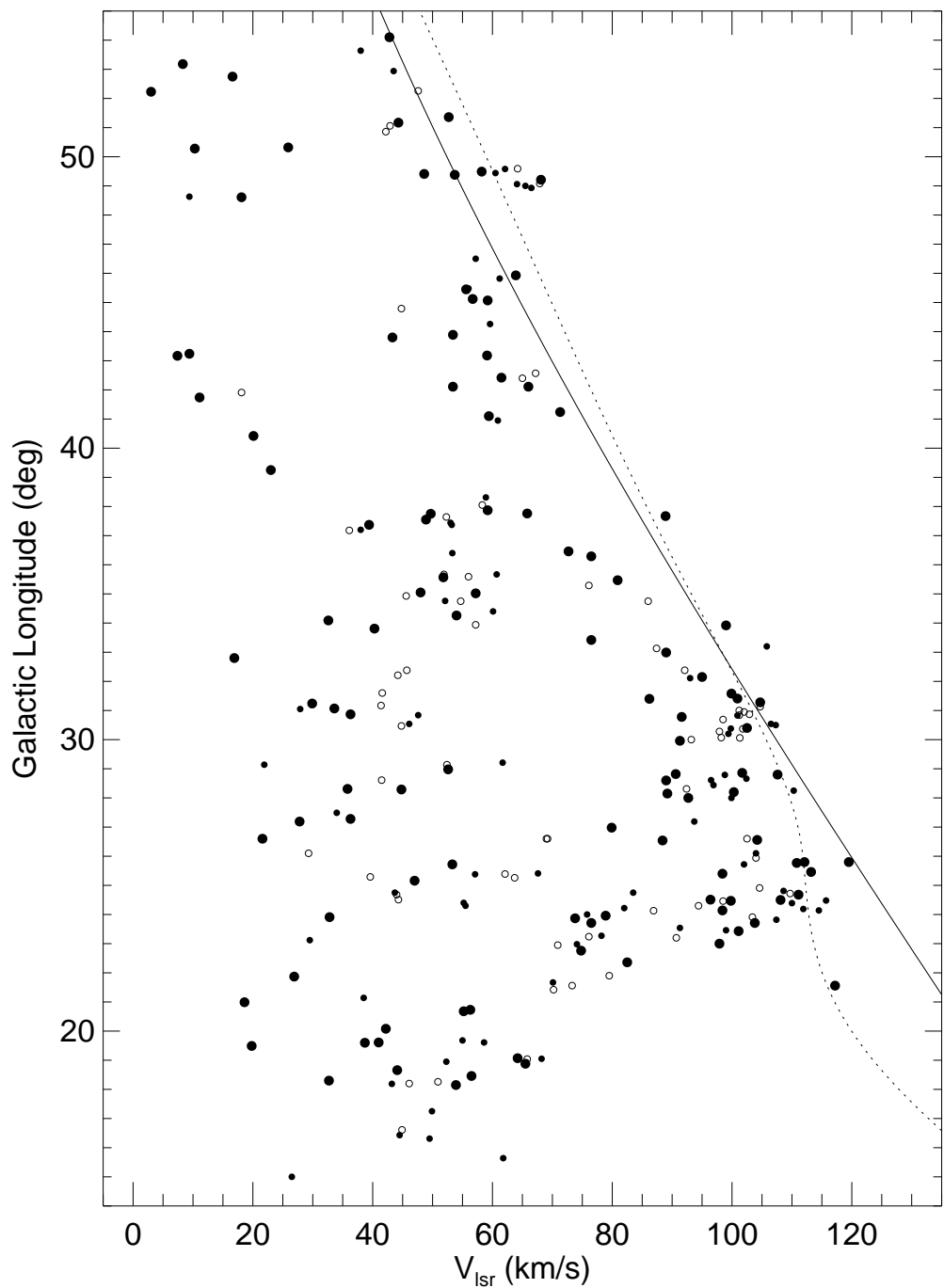}
\caption{Longitude-LSR velocity diagram for 253 \hii\ regions in our
sample that show associated \cor\ emission.  Symbols indicate the
confidence parameter of each nebula: $CP$ A (large filled circles), B
(small filled circles), or C (small open circles).  The lines show the
locii of the LSR terminal velocity expected from two different
Galactic rotation curve models: \citet{clemens85} (dotted line) and
\citet{brand86} (full line).  }

\label{fig:co_lv}
\end{figure}

Here we focus on the 253 nebulae with positive CO/\hii\ associations,
the subset of our \hii\ region sample with confidence parameter values
of A, B or C.  The $(l,V)$ distribution of these nebulae is shown in
Figure \ref{fig:co_lv} for $CP$ A (large filled circles), B (small
filled circles), and C (small open circles) sources.  Fully 90\% of UC
and compact nebulae have $CP$ A, B, or C quality CO/\hii\ associations
whereas only 64\% of the diffuse nebulae do.  Diffuse \hii\ regions
are probably older on average than either UC or compact nebulae so
their significantly lower associate rate with molecular gas provides
the first hint of evolutionary effects in our sample.

Table \ref{tab:groups} summarizes the mean properties we derive here
for this sample. Listed are the mean and standard deviation ($1
\sigma$) for each quantity.  This information is given for the entire
sample and also for various subsets of it: UC, compact, and diffuse
\hii\ regions as well as sources of $CP$ A, B and C.
Table \ref{tab:groups} lists the number of \hii\ regions in each
category, the absolute value of the velocity difference between the
\cor\, molecular clump and the RRL $V_{\rm LSR}$, the line intensity (main beam
brightness temperature), the FWHM line width, the size of the
associated molecular clump, the CO excitation temperature and the
\cor\, column density.

We estimate the average \cor\ column density towards our sources  
using the \citet{rohlfs} analysis:

\begin{equation} \label{eq:13cocolumn} 
N(^{13} {\rm CO})[{\rm cm^{-2}}] = 2.42 \times 10^{14} \frac{T_{\rm ex} 
\int \tau_{13}\, dv}{1 - {\rm exp}(-5.29)\,/\,T_{\rm ex}} 
\end{equation}

\noindent where \tex\, is the excitation temperature and $\tau_{13}$
is the optical depth of the \cor\, line.  We assume the \cor\ emission
is optically thin and use the Gaussian fit \cor\ line parameters to
find the optical depth integral in Eq.\ref{eq:13cocolumn}.  Both
the optical depth and Equation \ref{eq:13cocolumn}, however, depend on
the excitation temperature, \tex.

We use the UMSB \co\ $J = 1
\rightarrow 0$ survey \citep{sanders86} to estimate \tex\ for each
source.  We assume \co\, is optically thick and use the radiative
transfer equation for the $J = 1 \rightarrow 0$ transition to
calculate the excitation temperature from the observed main beam
brightness temperature, $T_{\rm MB}^{12}$, of the \co\, line:

\begin{equation} \label{eq:tex} 
T_{\rm ex} = 5.5 \left/ \ln \left( 1+
\frac{5.5}{T_{\rm MB}^{12}+0.82} \right) \right..
\end{equation} 

\noindent Eq. \ref{eq:tex} holds as long as: (1) the \co\, and
\cor\, emitting gas is in LTE at the same excitation temperature; (2)
this gas fills the same volume without clumping; and (3) there are no
background continuum sources.  (For our nebulae the \hii\ region
continuum is subtracted when the spectral baselines are removed.)

For each nebula we first compute an average spectrum from the UMSB
survey datacube in exactly the same way as we did for the GRS data.
We use the same ($l,b$) positions from the identical threshold
selected region for this average.  We then fit Gaussians to these
average spectra. If we fit multiple Gaussians to the GRS data, we
attempt to fit the same components to the UMSB data. The spectral
resolution of the UMSB survey is 1\kms, so this was not always
possible as lines resolved in the GRS are blended in the UMSB survey.

Because the pointings of the UMSB survey are further apart than in the
GRS ($180 \arcsec$ compared to $22 \arcsec$), the UMSB survey
underestimates the \co\ emission for the small molecular clumps found
in the GRS. For a given velocity, each pixel in the UMSB survey
represents $\sim 70$ pixels in the GRS. Each threshold selected region
contains $\sim 20$ GRS pixels on average, so for the majority of our
sources we use only one UMSB pointing to estimate the excitation
temperature appropriate for the \co\ emission.  Furthermore, this
pointing can be as far away as $\sim 120\arcsec$ from the GRS
position.

\begin{figure}
\epsscale{1.0}
\plotone{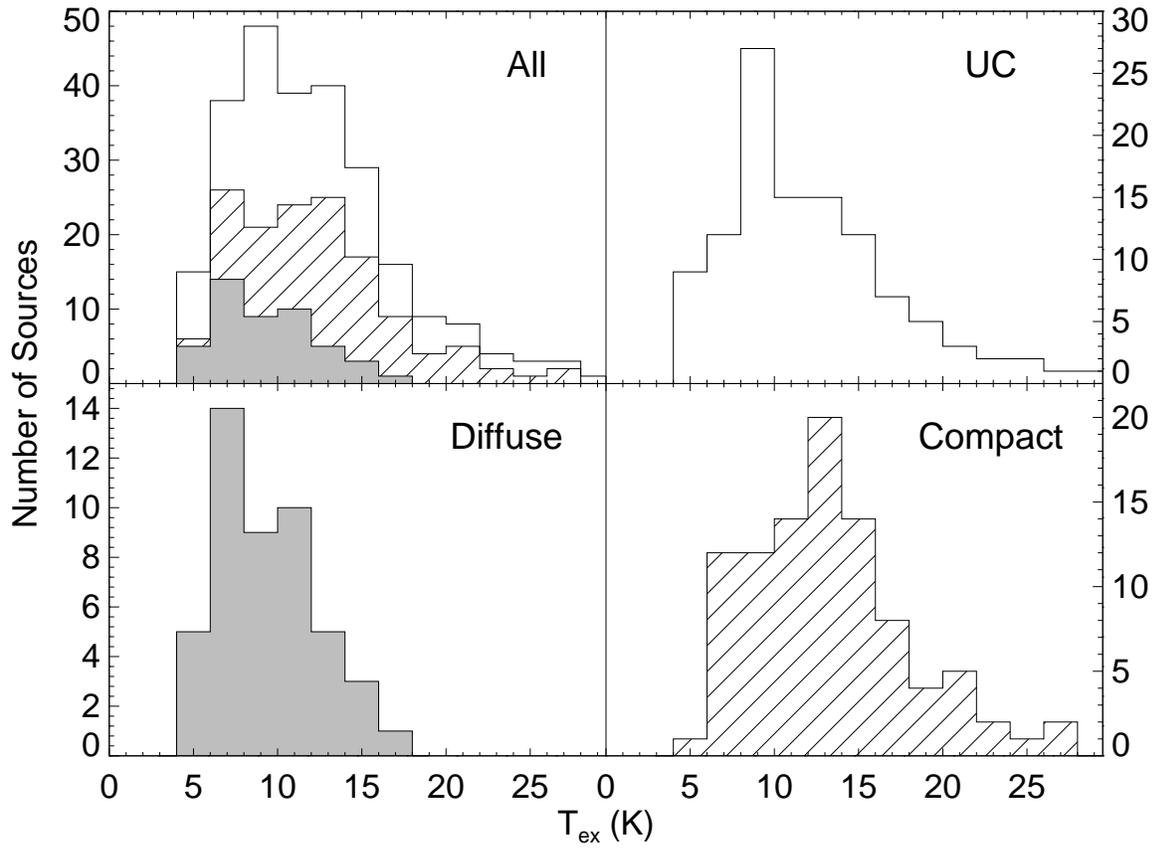}

\caption{Nebular excitation temperature, \tex, derived from the \co\, survey
of \citet{sanders86}. Panels are the same as in Figure
\ref{fig:confidence}.}

\label{fig:t_ex}
\end{figure}

The distribution of excitation temperatures we derive is shown in
Figure \ref{fig:t_ex}.  All \hii\ regions in our sample have very
similar excitation temperatures near the standard 10\,K value assumed
for molecular clouds.  We expected the excitation temperature of the
UC regions in particular to be higher than this standard value as the
CO gas is nearer to the exciting star. That we do not see hotter
temperatures associated with younger regions is probably due to the
effect of the undersampling of the \co\ emission by the UMSB survey.
The mean \co\ to \cor\ ratio for molecular clumps smaller than the
beam spacing of the UMSB survey, $3\arcmin$, is $\sim2$ while this
ratio is $\sim 3$ for molecular clumps larger than $3\arcmin$.  Our
excitation temperatures, and hence column densities, are therefore
lower limits.  Because of their small size, UCs are affected more by
the difference in sampling between the GRS and the UMSB survey.  The
UCs also suffer from beam dilution which will lower the inferred
excitation temperature.

\begin{figure}
\epsscale{1.0}
\plotone{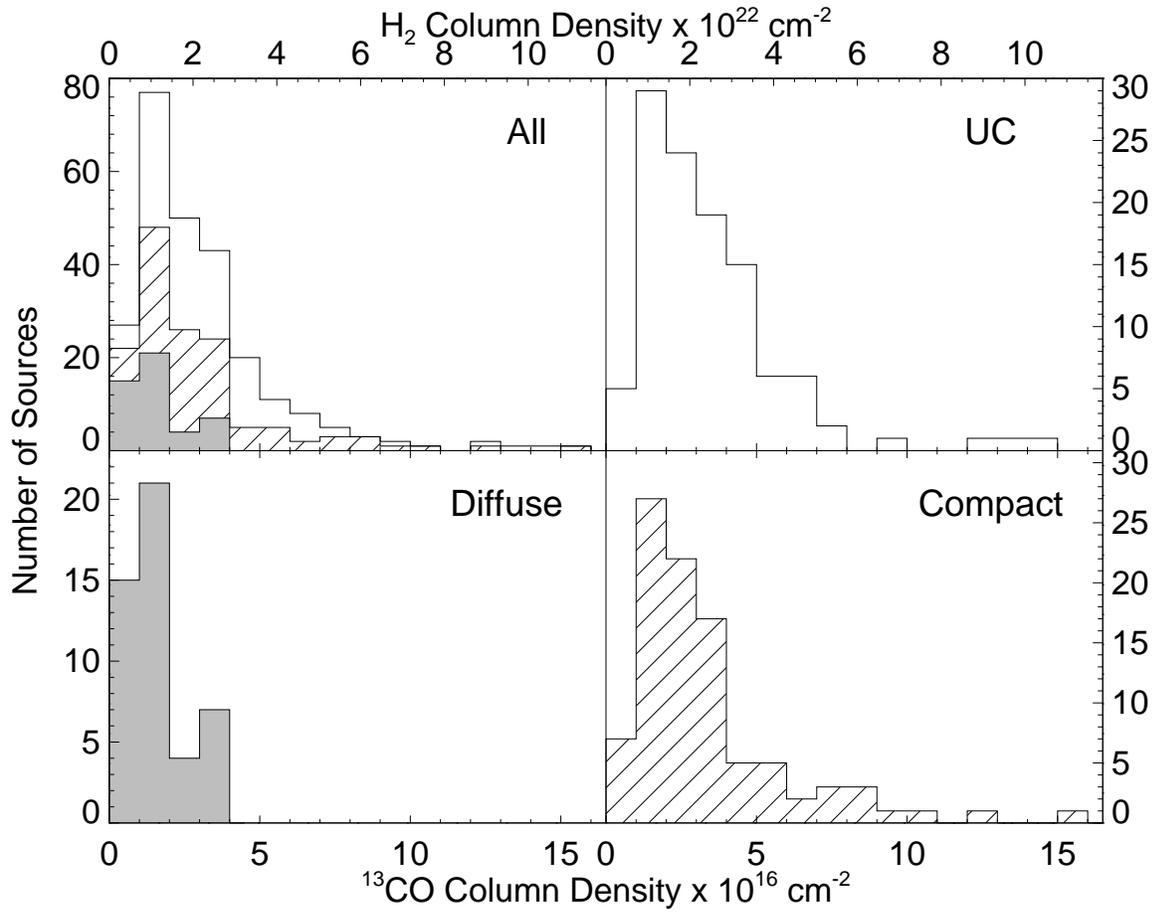}

\caption{Estimated H$_2$ column density (top axis) and \cor\, column
density (bottom axis) for the molecular clumps associated with our \hii\
regions. Panels are the same as in Figure \ref{fig:confidence}.}

\label{fig:column}
\end{figure}

Nevertheless, we use these excitation temperatures to compute the
\cor\, column density for each source using Eq. \ref{eq:13cocolumn}.
These column densities are listed in Table \ref{tab:diffuse}.  We then
estimate the $\rm{H}_2$ column density,

\begin{equation} \label{eq:h2column} 
N({\rm H_2}) = \left[\frac{^{12} 
{\rm CO}}{^{13} {\rm CO}} \right] \times \left[\frac{\rm H_2}{^{12} {\rm
CO}} \right] \times N(^{13} {\rm CO}), 
\end{equation} 

\noindent by assuming constant values for these abundance ratios.
Following \citet{simon01} we adopt a \co/\cor\ ratio of 45 and a $\rm
H_2$/\co\ ratio of \nexpo{8}{-5}.  Our \cor\, and $\rm H_2$ column
density results are summarized in Figure \ref{fig:column}.  As
expected, the UC nebulae have on average the highest column densities
and the diffuse nebulae the lowest.  Too, the $CP$ A sources have on
average over twice the column density of the $CP$ C nebulae.

\begin{figure}
\epsscale{1.0}
\plotone{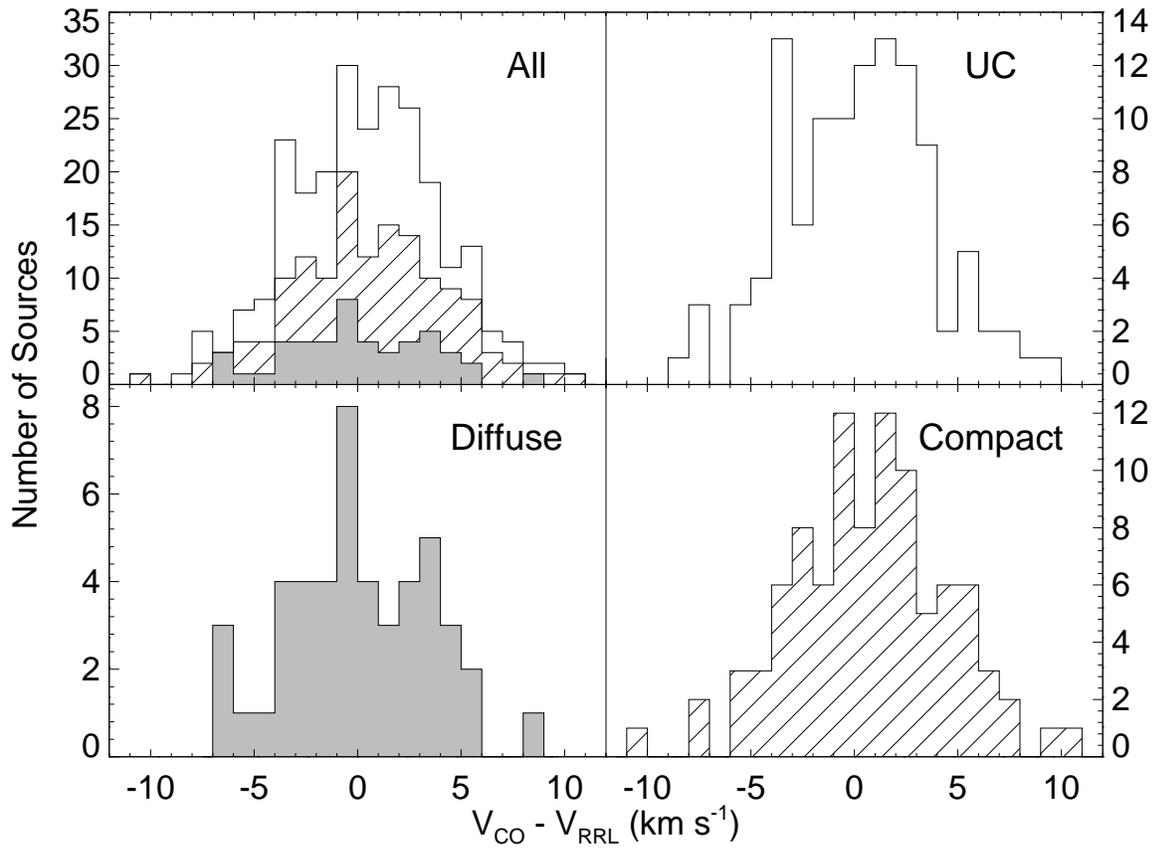}

\caption{The LSR velocity difference between the \hii\ region RRL and
molecular gas \cor\, velocity. The distribution peaks near zero, as
expected. Panels are the same as in Figure \ref{fig:confidence}.}

\label{fig:vel_diff_by_code}
\end{figure}

We find that the CO gas has on average only a small velocity offset
from the \hii\ region RRL velocity. Figure \ref{fig:vel_diff_by_code}
shows the difference between the velocity of the CO gas and the RRL
velocity.  There is no difference in velocity offset between the
various types of \hii\ regions.  The Gaussian fit to the entire
distribution is centered at 0.4 \kms\ with a FWHM of 8.5 \kms,
whereas the mean of the distribution is $0.2\pm 3.8$ \kms. The fact
that the distribution is centered at zero velocity offset is to be
expected for a RRL selected sample of \hii\ regions and an optically
thin tracer such as \cor.

This result is in contrast to optically selected samples where a 
positive velocity offset was found \citep[e.g., ][]{fich90}.  Optical
samples choose specific CO/\hii\ region line of sight geometries.
Face-on or edge-on blister sources such as the Orion nebula and M17
should dominate these samples.  In our sample there is no preferred
radial location for \hii\ regions within molecular clouds.  Because we
do not know the CO/\hii\ geometry with respect to the line of sight,
the absolute value of the CO/RRL velocity difference will be a direct
measure of any systematic velocity offset between the molecular and
ionized gas.  Table \ref{tab:groups} therefore lists the mean absolute
value of the CO/RRL velocity difference; it shows that there is a
$\sim\,3 \kms$ average flow velocity between the molecular and ionized
gas for the nebulae in our sample.  This value is independent of the
type of \hii\ region, but increases slightly as the $CP$ value decreases from A to C.

\begin{figure}
\epsscale{1.0}
\plotone{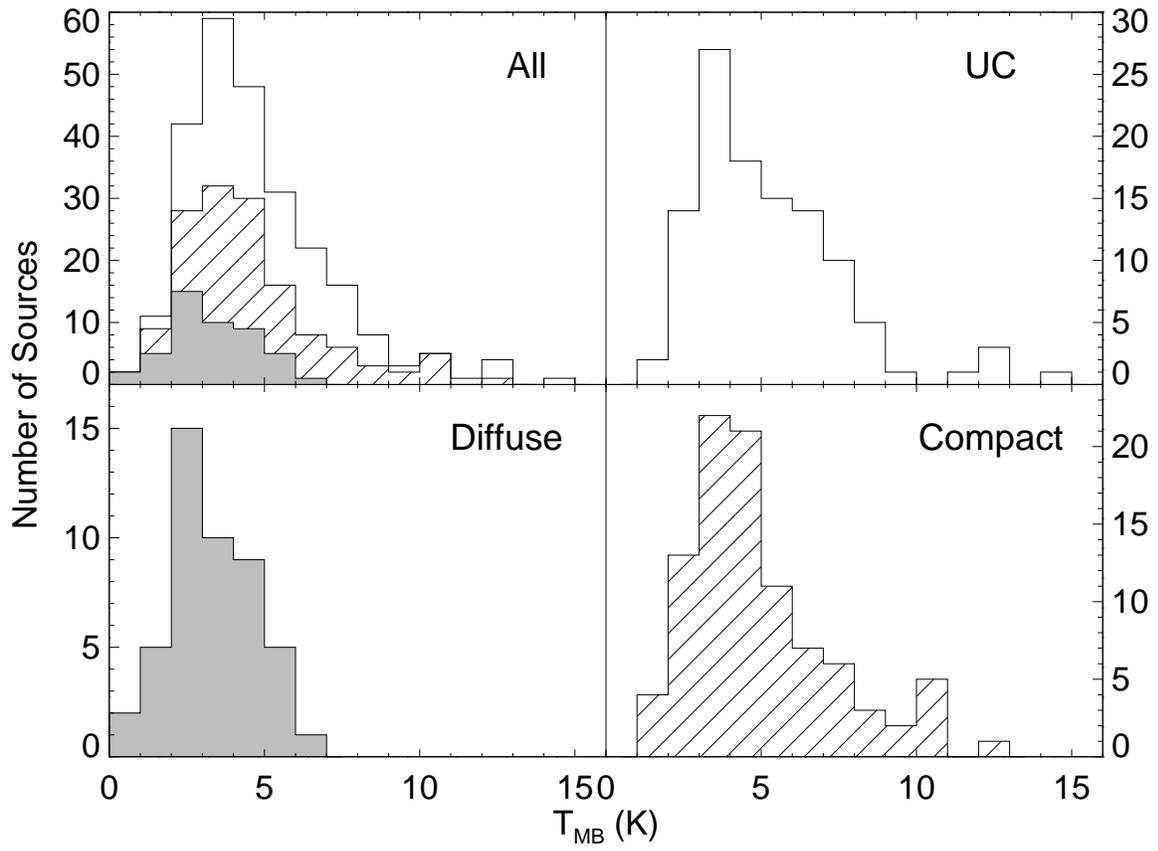}

\caption{Main beam brightness temperature \cor\, line intensity from Gaussian fits. 
Panels are the same as in Figure \ref{fig:confidence}.}

\label{fig:intensity}
\end{figure}

Figure \ref{fig:intensity} shows the distribution of \cor\ line
intensities for our sample. Based on the evolutionary model of \hii\
regions the natal cloud is gradually dissipated by photo-dissociation,
photo-ionization, and expanding motions.  We expect the \cor\ density
to decrease as the region progresses from UC to compact and then,
finally to diffuse.  Assuming \cor\, is optically thin, or at least
marginally so, the higher densities found in UC regions would lead to
higher line intensities, whereas compact sources should show lower
line intensities, and diffuse sources the lowest. This hypothesis is
only partially borne out as UC and compact regions share the same
distribution, averaging $5.21 \pm 0.23 \K$ and $4.96 \pm 0.24 \K$,
respectively. Diffuse regions do show lower line intensities of $3.32
\pm 0.19$. The errors quoted here are the standard errors of the 
mean, s.e.m. $\equiv \sigma/\sqrt{N}$.

\begin{figure}
\epsscale{1.0}
\plotone{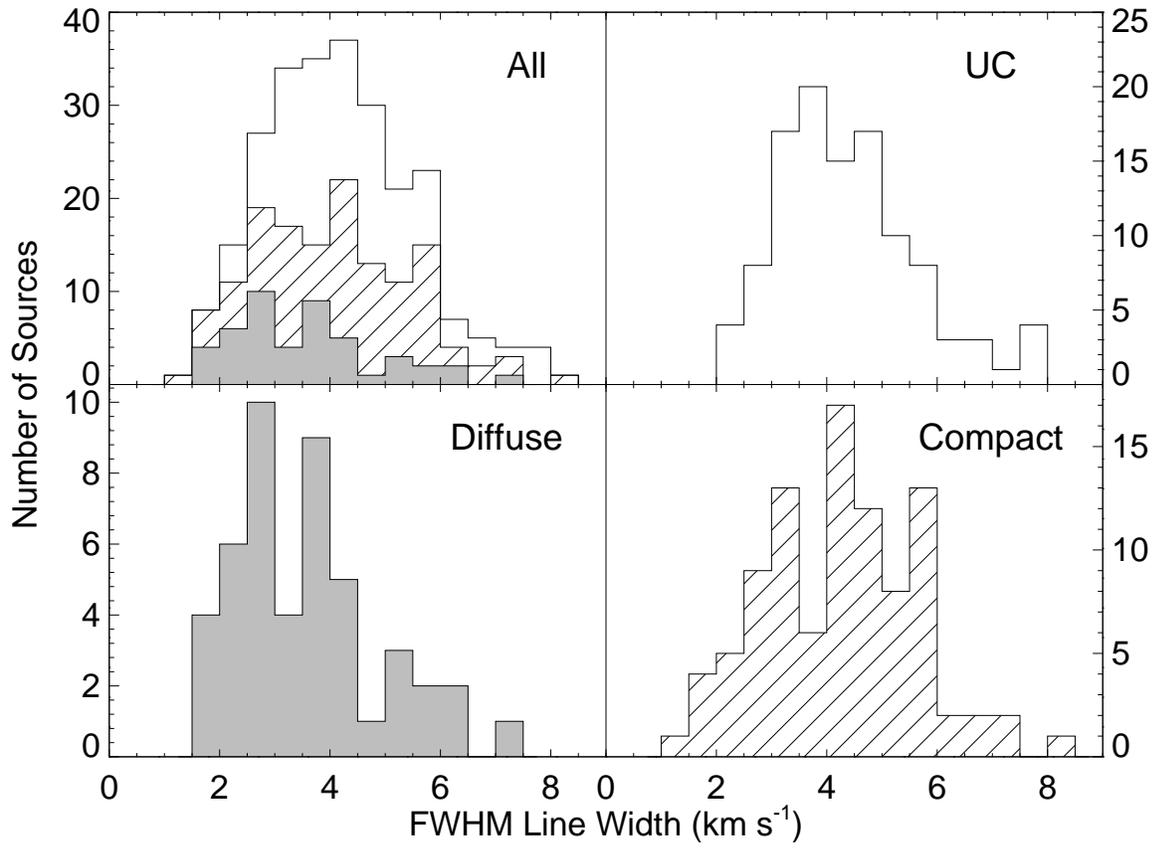}

\caption{\cor\, FWHM line width from Gaussian fits.  
Panels are the same as in Figure \ref{fig:confidence}.}

\label{fig:linewidth}
\end{figure}

Figure \ref{fig:linewidth} shows the distribution of \cor\ line widths
for our sample. The Gaussian fit to this distribution is centered at
4.0 \kms\ with a FWHM of 3.2 \kms. The mean of this distribution is
$4.20\pm0.09$ \kms. We expected UC \hii\ regions to have significantly
broader lines than compact \hii\ regions because the molecular gas is
closer to the exciting star and the outflows should be stronger.  Once
again, this is not the case: the UC and compact distributions are very
similar, averaging $4.43\pm0.13\kms$ (s.e.m) and $4.23 \pm 0.14\kms$
(s.e.m.), respectively.  UC and compact regions do have broader lines
than the diffuse regions which average $3.56\pm0.20 \kms$ (s.e.m.).
This suggests that the central star(s) may no longer be significantly
heating molecular gas near diffuse \hii\ regions. 

This distribution of line widths is comparable to that found by
\citet{russeil04} in their single pointing survey of southern \hii\
regions. They find that the \cor\, $J = 1 \rightarrow 0$ line has an
average line width of 3.7 \kms\ with a standard deviation of 1.9
\kms.  In a study of UC \hii\ regions, \citet{kim03} find an average line
width of 6.8 \kms\ in the \cor\, $J=1 \rightarrow 0$ transition.  Their
calculation of line width, however, was based on the average spectrum
over the entire map area, which was as large as $30 \arcmin \times 40
\arcmin$. For the 8 UC regions in our sample that \citet{kim03} study,  
we measure a line width of 5.0 \kms\ whereas they measure 7.2
\kms.  Using the same large areas to compute the line widths for these
sources, we find an average line width of 7.3 \kms.

\begin{figure}
\epsscale{1.0}
\plotone{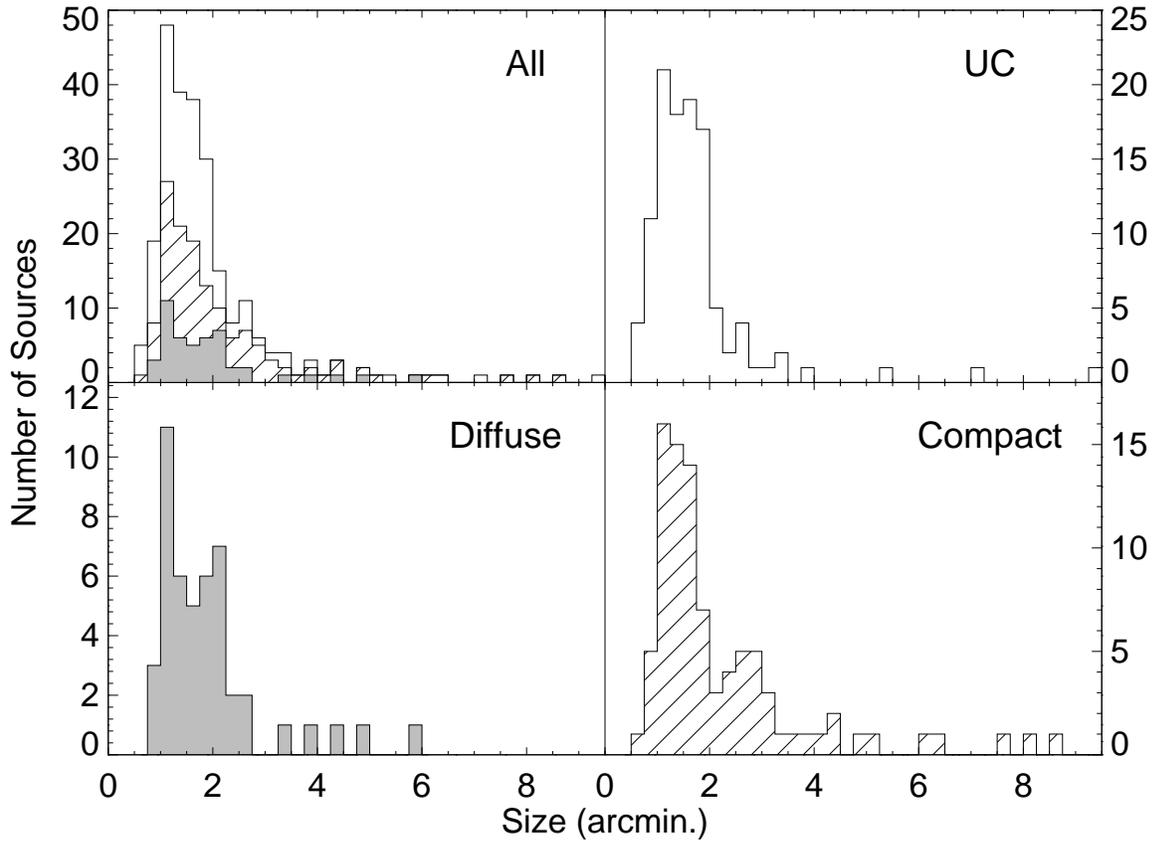}

\caption{Angular size defined as the geometric mean diameter of the fitted 
ellipses. Panels are the same as in Figure \ref{fig:confidence}.}

\label{fig:angsize}
\end{figure}

The angular size distribution of the \hii\/CO sources is shown in
Figure \ref{fig:angsize}. These sizes are defined as the geometric
mean of the major ($2 \times a$) and minor ($2 \times b$) axes of the
fitted ellipse, $2\sqrt{ab}$. (See \S\ref{sec:singlepointing} for our
ellipse fitting procedure.)  There are 6 sources that have sizes
greater than $10 \arcmin$ that are not plotted in Figure
\ref{fig:angsize}. These sources are invariably clumps in a
large region of extended molecular emission, which makes our method of
determining the angular size unreliable.

The molecular clumps associated with UC regions do show the smallest
sizes, as expected, averaging $1 \farcm 7 \pm 0 \farcm 1$ (s.e.m.). Compact
\hii\ regions are slightly larger, averaging $2 \farcm 2 \pm 0 \farcm
1$ (s.e.m.).  The average size of the molecular gas associated with diffuse
\hii\ regions lies in between that of UC and compact \hii\ regions,
averaging $1 \farcm 9 \pm 0 \farcm 1$ (s.e.m.).  We have removed the 6
sources with sizes greater than $10\arcmin$ from the statistical
analysis. The molecular gas around many diffuse regions is fragmented,
which leads to the small angular sizes we measure.  
Since we only associate a single molecular clump of contiguous pixels
with each nebula, for diffuse \hii\ regions we probably have not
characterized all the associated molecular gas.


\subsection{Comparison with GRS Molecular Clumps}\label{sec:GRS}

The properties of molecular clumps in the GRS were analyzed down to
size scales of $\sim 1\arcmin$ (Rathborne et al. 2008, in
preparation).  The contiguous pixel finding algorithm {\it CLUMPFIND}
\citep{williams94} was used to locate Giant Molecular Clouds (GMCs)
within the GRS.  Then, by altering the size threshold in {\it
CLUMPFIND}, the clumps within the GMCs were identified and
characterized.  The distribution of peak intensities and line widths
for these clumps shows a Gaussian core with an exponential tail at
high values of each parameter.  The break points where the
distributions turn over from being dominated by the Gaussian core to
being dominated by the exponential tail are roughly 4 K and 2 \kms.
By number the vast majority of these GRS molecular clumps have line
intensities below 4 K and line widths below 2 \kms.

\begin{figure}
\epsscale{1.0}
\plotone{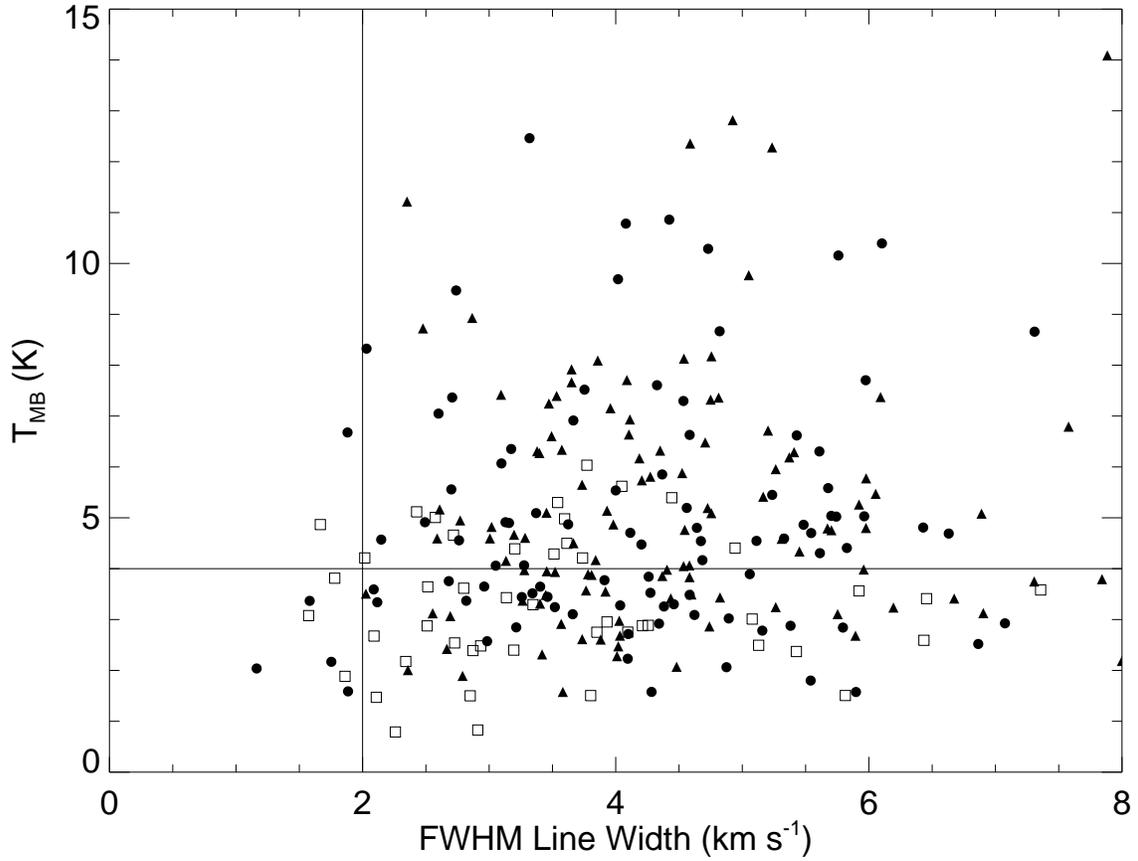}

\caption{\cor\, line intensity plotted as a function of FWHM line
width.  Shown are all the $CP$ A, B, or C sources: UC (filled triangles),
Compact (filled circles), and Diffuse (open squares) nebulae.  The lines
dividing the figure into quadrants are from an analysis of molecular
clumps in the GRS dataset (see \S \ref{sec:GRS}).}

\label{fig:t_vs_fwhm}

\end{figure}

The molecular gas associated with \hii\ regions has on average a
greater line intensity and larger line width compared to molecular
clumps in the GRS.  We plot in Figure \ref{fig:t_vs_fwhm} the line
intensity verses the FWHM line width for the molecular clumps
associated with our \hii\ regions: UC (filled triangles), Compact (filled circles),
and Diffuse (open squares) nebulae. The solid lines divide the plot
into quadrants according to the break points of the GRS clumps.  A
similar plot was used by \citet{clemens88} to show that the small
clouds in their optically selected molecular cloud sample were cool
and quiescent.

The lower left quadrant in Figure \ref{fig:t_vs_fwhm} should be
populated by cold quiescent clouds.  These objects are neither making
stars nor being externally heated.  The upper left quadrant should
contain a population of clumps that are heated externally.  These
clouds are warm (or have high column densities), but do not have the
non-thermal motions that would be present if they posessed a central
star.  The upper right quadrant should contain molecular gas
associated with embedded massive stars.  The lower right quadrant
should contain embedded protostars.  These large line width objects
are likely active sites of star formation, or near an active site.
Thus the lower, $<4\K$, part of the plot has a pre-stellar population
whereas the upper part contains clouds that are affected by local
massive stars.

The UC and compact nebulae occupy the same region of Figure
\ref{fig:t_vs_fwhm}; they have similar molecular properties.  The
diffuse regions, however, have lower line intensities and line widths;
they are similar to the general population of molecular clumps. These
clumps are no longer being heated significantly by the ionizing star.

Most of our sources with associated \cor, 54\%, lie in the upper right
quadrant of Figure \ref{fig:t_vs_fwhm} where large line intensities
and broad line widths suggest active star formation.  The vast
majority of GRS clumps, as well as most of the molecular clouds in
\citet{clemens88}, reside in the lower left quadrant.  The bulk of our
remaining nebulae, 42\%, lie in the lower right quadrant.  The UC,
compact, and diffuse regions all have a significant population in this
quadrant.  It is tempting to think that these low line intensities are
due to the decreased optical depth of \cor\, but Figure
\ref{fig:t_vs_fwhm} looks very similar to the same plot in
\citet{russeil04} made using the optically thick \co\, $J=2
\rightarrow 1$ transition.  

\subsection{Are Molecular Cloud/H~{\bf\small II} Region Associations Real?
\label{sec:reality}}

Are these CO/\hii\ region associations really sources that are having
a direct physical interaction between the \hii\ region and ambient
molecular gas?  The associations are based on morphological matches in
position and velocity between the \cor\ gas, RRL velocity, and radio
continuum emission.  But correlation does not imply causality: these
matches could in principle be a coincidental juxtaposition projected
on the sky into the same solid angle by molecular clouds and \hii\
regions located at entirely different places along the line of sight.

That we require a morphological match in $(l, b, V)$ -- space places a
severe constraint on a false positive association.  Mere $(l, b)$
coincidence is not enough; the velocity also needs to match.  The
kinematic distance ambiguity in the Inner Galaxy makes it possible for
the \hii\ region and CO cloud to be at different line of sight
positions despite having nearly identical radial velocities.  But
these are special places because only they share the same LSR
velocity.  Assessing the quantitative probability of a false positive
association is beyond the scope of this paper.  To our knowledge no
one has yet done the detailed modelling this would require.  One needs
to know the Galactic distribution of the clouds which posits a
detailed knowledge of Galactic structure.  For a false positive
association we require that there not be a cloud at the \hii\ region
position, but that there be a cloud at the other kinematic distance.
One thus needs to evaluate separately for each \hii\ region the line
of sight distance derivative of the LSR velocity, $dV/dr$ in order to
assess the path lengths at the near and far kinematic distances that
must be populated.  This requires a detailed knowledge of Galactic
kinematics, including streaming motions caused by spiral arms.  With
this information one might be able to estimate the probability of a
false positive association.  We probably do not know enough about
either Galactic structure or Galactic kinematics to do this.

%
%

The fact, however, that $\sim\,20\%$ of the \hii\ regions do not have
associated CO gas (\S \ref{sec:properties}) is evidence that suggests
chance line of sight superpositions in $(l,b,V)$ -- space are rare.
Furthermore, Figure \ref{fig:t_vs_fwhm} provides strong support for
the physical reality of our CO/\hii\ region associations.  Our \cor\
clouds are not only near to the \hii\ regions in $(l,b,V)$ -- space,
but their spectra also have the trademarks of star formation: bright
lines and large line widths.  This is in marked contrast to the
spectral line properties of the vast majority of GRS clouds (see \S
\ref{sec:GRS}).  Only 1/3 of the GRS clumps have peak intensities $> 4
\K$ whereas 55\% of our CO/\hii\ associated clouds do.  Only 1/6 of the
GRS clouds have line widths $> 2 \kms$; nearly all of our clouds,
96\%, exceed this value.
We conclude that most of the CO/\hii\ region associations must be
nebulae with real physical interactions between the molecular and
ionized gas.

\section{The GRS H\,{\small \bf II} Region Catalog}\label{sec:catalog}

Our analysis here produced a catalog of Figure \ref{fig:images_a} type
images and physical properties for the sample of 301 Galactic \hii\
regions.  We created a
website\footnote[8]{http://www.bu.edu/iar/hii\_regions} to
give everyone access to this information. In addition to the images of
nebular \wco, this website has the average
\cor\ spectrum of each source as well as all the information found in
Tables \ref{tab:diffuse} and \ref{tab:properties}.  We expect this
website to be an evolving database compiling additional information
about these nebulae as it becomes available.

\section{Future Work}\label{sec:future}

Our sample of CO/\hii\ region associations will enable many further
studies of the properties of star forming regions at all stages of
their evolution.  The most important parameter that is missing here is
the distance to each nebula.  Knowing the distance would enable us to
derive the intrinsic physical properties of each nebula, establishing
their physical sizes and turning column densities and line intensities
into masses and luminosities.

\citet{anderson08}[AB hereafter] use \hi\ absorption studies to 
derive kinematic distances toward all the \hii\ regions with
associated molecular gas.  All our nebulae are in the first Galactic
quadrant, so their distances are degenerate due to the kinematic
distance ambiguity.  Using the fact that \hi\ absorbs thermal
continuum from the \hii\ region, AB use the VGPS 21cm \hi\
emission line maps to remove this degeneracy \citep[see][]{kuchar94}.
This is a proven technique as there is sufficient residual cold \hi\
associated with almost all GRS molecular clouds to produce significant
absorption \citep{jackson02, kolpak03, flynn04}.  We shall then use
these distances to analyze this nebular sample and derive the physical
properties of the dust and gas (ionized, atomic \& molecular).  The
completion of the {\it Spitzer} GLIMPSE \citep{benjamin03} and MIPSGAL
(Carey et al. 2008 in preperation) surveys, together with the GRS
\citep{jackson06}, MAGPIS \citep{helfand06}, NVSS \citep{condon98} 
and the VGPS surveys enable for the first time a multi-wavelength
analysis of the physical properties and evolutionary state of a large
sample of inner Galaxy \hii\ regions.  Due to our large sample size,
we will be able to find examples of \hii\ regions at all evolutionary
stages.

\section{Summary}

We analyzed the GRS \cor\, molecular gas associated with all known \hii\
regions covered by the GRS using multiple analysis techniques. Our
sample includes 301 regions: 123 UC, 105 compact and 73 diffuse \hii\
regions. We found that 80\% of our \hii\ regions showed positive
molecular associations, with UCs having the highest association
percentage and diffuse regions the lowest. About 5\% of our sample
showed no molecular emission whatsoever. We hypothesize that some of
these non-detections represent an older population of \hii\ regions
where the molecular gas has been displaced from the central star or
stars.
We found that the molecular properties of UC and compact \hii\ regions
are quite similar, with line widths averaging $\sim 4 \kms$ and \cor\,
column densities of about $3.5 \times 10^{16} {\rm \,cm^{-2}}$. The
molecular gas associated with diffuse regions has properties more
consistent with quiescent clouds. The molecular gas properties of our
sample nebulae are consistent with an evolutionary sequence wherein
small, dense molecular gas clumps associated with UC \hii\ regions
grow into older compact nebulae and finally fragment and dissipate
into large, diffuse nebulae.

\acknowledgments
This publication makes use of molecular line data from the Boston
University-FCRAO Galactic Ring Survey (GRS). The GRS is a joint
project of Boston University and Five College Radio Astronomy
Observatory, funded by the National Science Foundation under grants
AST-9800334, AST-0098562, \& AST-0100793.
\nraoblurb


\clearpage

\begin{deluxetable}{lcc}
\tabletypesize{\scriptsize}
\tablecaption{Faux and Anomalous Sources}
\tablewidth{0pt}
\tablehead{
\colhead{Source} & 
\colhead{Notes} & 
\colhead{Reference}
}

\startdata

C14.32$+$0.13 & SNR & a \\
D15.45$+$0.19 & SNR & a \\
D15.52$-$0.14 & SNR & b \\
U16.58$-$0.05 & No continuum peak &  \\
D17.23$+$0.39 & Probably an evolved star &  c, d \\
U17.64$+$0.15 & No continuum peak &  \\
C18.64$-$0.29 & SNR & a, e \\
U19.12$-$0.34 & No continuum peak &  \\ 
U19.36$-$0.02 & No continuum peak &  \\
C19.88$-$0.53 & No continuum peak &  \\
C20.26$-$0.89 & No IR &  \\
C20.48$+$0.17 & SNR & a, e \\
D21.56$-$0.11 & SNR & a, e  \\
D22.04$+$0.05 & No continuum peak &  \\
D2216$-$0.16 & Star & c \\
D22.40$-$0.37 & SNR & e \\
C22.94$-$0.07 & No IR &  \\ 
C23.07$-$0.37 & No IR  - Part of SNR? &  \\
C23.07$-$0.25 & No IR  - Part of SNR? &  \\
D23.16$+$0.02 & No continuum peak &  \\
U23.24$-$0.24 & No continuum peak &  \\
D26.47$+$0.02 & WR or LBV & f \\
U26.51$+$0.28 & No continuum peak &  \\
C27.13$-$0.00 & SNR & e \\
C29.09$-$0.71 & SNR & e \\
D29.55$+$0.11 & SNR & g \\
U30.42$+$0.46 & No continuum peak &  \\
D30.69$-$0.63 & No IR &  \\
U30.82$+$0.27 & No continuum peak &  \\ 
C30.85$+$0.13 & SNR & e \\
C31.05$+$0.48 & SNR & e \\
D31.61$+$0.33 & SNR & e \\
D31.82$-$0.12 & SNR & e \\
U33.24$+$0.01 & No continuum peak &  \\
C45.48$+$0.18 & No continuum peak &  \\ 
U49.67$-$0.45 & No continuum peak &  \\
C50.23$+$0.33 & No IR &  \\
U53.63$+$0.02 & No continuum peak &  \\

\enddata
\tablenotetext{a}{\citet{brogan06}}
\tablenotetext{b}{\citet{brogan06} (low confidence detection)}
\tablenotetext{c}{\citet{stephenson92}}
\tablenotetext{d}{\citet{kwok97}}
\tablenotetext{e}{\citet{helfand06}}
\tablenotetext{f}{\citet{clark03}}
\tablenotetext{g}{\citet{gaensler99}}
\label{tab:anomalous}
\end{deluxetable}

\begin{deluxetable}{lcccccll}
\tabletypesize{\scriptsize}
\tablecaption{\hii\ Region Source Sample}
\tablewidth{0pt}
\tablehead{
\colhead{Source} & 
\colhead{l} & 
\colhead{b} & 
\colhead{RA(2000.0)} & 
\colhead{DEC(2000.0)} & 
\colhead{$V_{\rm LSR}$} &
\colhead{Reference} &
\colhead{Comments} \\ 

\colhead{} & 
\colhead{($\degr$)} & 
\colhead{($\degr$)} & 
\colhead{(h m s)} & 
\colhead{($\degr$ $\arcmin$ $\arcsec$)} & 
\colhead{(\kms)}  
}
\startdata
D15.00$+$0.05a & 15.00 & $+$0.05 & 18 17 41 & $-$15 52 50 &  \phn \phn 26.5 $\pm$ 1.6 &  L96 &   \\
D15.00$+$0.05b & \nodata & \nodata &  \nodata  &  \nodata  &  \phn \phn 63.5 $\pm$ 1.7 &  L96 &   \\
D15.64$-$0.24 & 15.64 & $-$0.24 & 18 19 60 & $-$15 27 10 &  \phn \phn 61.8 $\pm$ 1.3 &  L96 &   \\
C16.31$-$0.16 & 16.31 & $-$0.16 & 18 21 01 & $-$14 49 30 &  \phn \phn 49.5 $\pm$ 0.7 &  L89 &   \\
C16.43$-$0.20 & 16.43 & $-$0.20 & 18 21 24 & $-$14 44 10 &  \phn \phn 44.5 $\pm$ 0.9 &  L89 &   \\
D16.61$-$0.32 & 16.61 & $-$0.32 & 18 22 11 & $-$14 38 10 &  \phn \phn 44.9 $\pm$ 0.6 &  L89 &  a \\
D16.89$+$0.13 & 16.89 & $+$0.13 & 18 21 05 & $-$14 10 30 &  \phn \phn 42.3 $\pm$ 1.6 &  L96 &   \\
D17.25$-$0.20a & 17.25 & $-$0.20 & 18 22 59 & $-$14 00 50 &  \phn \phn 49.9 $\pm$ 1.4 &  L96 &   \\
D17.25$-$0.20b & \nodata & \nodata &  \nodata  &  \nodata  &  \phn \phn 96.5 $\pm$ 1.9 &  L96 &   \\
U18.15$-$0.28 & 18.15 & $-$0.28 & 18 25 01 & $-$13 15 20 &  \phn \phn 53.9 $\pm$ 0.4 &  L89 &   \\

\enddata

\tablenotetext{a}{Changed nebular classification from compact}

\tablecomments{Table \ref{tab:diffuse} is published in its entirety in the electronic edition of the 
Astrophysical Journal Supplement Series.  A portion is shown here for
guidance regarding its form and content.}


\label{tab:diffuse}
\end{deluxetable}
 
\begin{deluxetable}{lcccccccccccc}
\tabletypesize{\scriptsize}
\rotate
\tablecaption{Properties of Molecular Cloud/\hii Region Sources}
\tablewidth{0pt}
\tablehead{
\colhead{} & 
\multicolumn{5}{c}{Fitted Ellipse Parameters} & 
\colhead{} &
\multicolumn{3}{c}{Fitted Gaussian Parameters} &
\colhead{} &
\colhead{} &
\colhead{} \\ \cline{2-6} \cline{8-10}

\colhead{Source} & 
\colhead{l} & 
\colhead{b} &
\colhead{Size} &
\colhead{Maj. $\times$ Min.} &
\colhead{PA} &
\colhead{} &
\colhead{V} & 
\colhead{$T_{\rm MB}$} & 
\colhead{$\Delta V$} &
\colhead{\tex} &
\colhead{$N(^{13}{\rm CO})$} &
\colhead{$CP$} \\

\colhead{} & 
\colhead{($\degr$)} & 
\colhead{($\degr$)} & 
\colhead{($\arcmin$)} & 
\colhead{($\arcmin \times \arcmin$)} & 
\colhead{($\degr$)} & 
\colhead{} &
\colhead{(\kms)} & 
\colhead{(K)} & 
\colhead{(\kms)} & 
\colhead{K} &
\colhead{($\times 10^{16} \rm \, cm^{-2}$)} &
\colhead{}
}
\startdata

D15.00$+$0.05a & 14.93 & $+$0.02 & \phn 2.1 & $\phn 1.6 \times  0.7\phn$ & +88.6 &   & \phn 25.85 $\pm$ 0.02 & \phn 6.03 $\pm$ 0.02 & \phn 3.77 $\pm$ 0.05 & 12.7 & \phn   3.2 & B \\
D15.00$+$0.05b & \nodata & \nodata & \nodata & \nodata & \nodata &   & \phn \nodata & \phn \nodata & \phn \nodata & \nodata & \nodata & E \\
D15.64$-$0.24 & 15.66 & $-$0.21 & \phn 2.0 & $\phn 2.1 \times  0.5\phn$ & +70.6 &   & \phn 56.96 $\pm$ 0.06 & \phn 1.88 $\pm$ 0.06 & \phn 1.86 $\pm$ 0.09 & \phn 8.6 & \phn   0.4 & B \\
C16.31$-$0.16 & 16.36 & $-$0.21 & \phn 1.7 & $\phn 0.9 \times  0.8\phn$ & +45.0 &   & \phn 47.56 $\pm$ 0.11 & \phn 4.17 $\pm$ 0.11 & \phn 4.68 $\pm$ 0.12 & 11.0 & \phn   2.6 & B \\
C16.43$-$0.20 & 16.36 & $-$0.21 & \phn 1.7 & $\phn 0.9 \times  0.8\phn$ & +14.9 &   & \phn 48.83 $\pm$ 0.02 & \phn 9.47 $\pm$ 0.02 & \phn 2.74 $\pm$ 0.04 & 13.9 & \phn   3.9 & B \\
D16.61$-$0.32 & 16.56 & $-$0.34 & \phn 4.8 & $\phn 2.9 \times  2.0\phn$ & +26.3 &   & \phn 43.23 $\pm$ 0.01 & \phn 5.39 $\pm$ 0.01 & \phn 4.44 $\pm$ 0.04 & 11.1 & \phn   3.2 & C \\
D16.89$+$0.13 & \nodata & \nodata & \nodata & \nodata & \nodata &   & \phn \nodata & \phn \nodata & \phn \nodata & \nodata & \nodata & D \\
D17.25$-$0.20a & 17.23 & $-$0.24 & \phn 3.9 & $\phn 2.7 \times  1.4\phn$ & $-$63.4 &   & \phn 44.57 $\pm$ 0.03 & \phn 5.62 $\pm$ 0.03 & \phn 4.05 $\pm$ 0.07 & 15.7 & \phn   3.6 & B \\
D17.25$-$0.20b & \nodata & \nodata & \nodata & \nodata & \nodata &   & \phn \nodata & \phn \nodata & \phn \nodata & \nodata & \nodata & D \\
U18.15$-$0.28 & 18.15 & $-$0.31 & \phn 2.2 & $\phn 1.7 \times  0.7\phn$ & +77.9 &   & \phn 52.53 $\pm$ 0.04 & \phn 7.32 $\pm$ 0.04 & \phn 4.75 $\pm$ 0.06 & 24.6 & \phn   7.6 & A \\

\enddata

\tablecomments{Table \ref{tab:properties} is published in its entirety in the electronic edition of the 
Astrophysical Journal Supplement Series.  A portion is shown here for
guidance regarding its form and content.}

\label{tab:properties}
\end{deluxetable}

\begin{deluxetable}{lccccccc}
\tabletypesize{\scriptsize}
\tablecaption{Mean Properties of Nebulae with Associated $^{13}{\rm CO}$}
\tablewidth{0pt}
\tablehead{
\colhead{} &
\colhead{N} &
\colhead{$|V|$ Offset} &
\colhead{$T_{\rm MB}$} &
\colhead{$\Delta V$} & 
\colhead{Size} &
\colhead{$T_{\rm ex}$} &
\colhead{$N({\rm ^{13} CO})$} \\
\colhead{} &
\colhead{} &
\colhead{(\kms)} & 
\colhead{(K)} &
\colhead{(\kms)} & 
\colhead{(\arcmin)} &
\colhead{(K)} &
\colhead{($\times 10^{16} {\,\rm cm}^{-2}$)} \\
}
\startdata

All & 253 &   2.98 $\pm$   2.41 &  4.77 $\pm$ 2.32 &   4.19 $\pm$  1.42 &   1.9
 $\pm$  1.3 &  12.1 $\pm$  4.8 &   3.1 $\pm$  2.6 \\
UC & 111 &   3.03 $\pm$   2.61 &  5.21 $\pm$ 2.40 &   4.43 $\pm$  1.38 &   1.7
 $\pm$  1.1 &  12.2 $\pm$  5.0 &   3.5 $\pm$  2.8 \\
Compact & \phn95 &   3.01 $\pm$   2.37 &  4.96 $\pm$ 2.36 &   4.23 $\pm$  1.40
 &   2.2 $\pm$  1.6 &  13.3 $\pm$  4.7 &   3.3 $\pm$  2.6 \\
Diffuse & \phn47 &   2.81 $\pm$   2.01 &  3.32 $\pm$ 1.29 &   3.56 $\pm$  1.37
 &   1.9 $\pm$  1.0 & \phn  9.3 $\pm$  3.0 &   1.6 $\pm$  1.0 \\
A & 112 &   2.65 $\pm$   2.06 &  5.63 $\pm$ 2.51 &   4.52 $\pm$  1.26 &   1.7
 $\pm$  0.8 &  12.8 $\pm$  5.1 &   4.0 $\pm$  3.0 \\
B & \phn 75 &   3.00 $\pm$   2.37 &  4.62 $\pm$ 2.06 &   4.05 $\pm$  1.31 & 
  2.1 $\pm$  1.5 &  12.1 $\pm$  4.4 &   2.8 $\pm$  2.0 \\
C & \phn 66 &   3.53 $\pm$   2.90 &  3.46 $\pm$ 1.51 &   3.80 $\pm$  1.66 & 
  2.2 $\pm$  1.7 &  11.0 $\pm$  4.4 &   1.9 $\pm$  1.9 \\

\enddata


\label{tab:groups}
\end{deluxetable}

\clearpage

\begin{thebibliography} {}



\bibitem[Altenhoff et al. (1979)] {altenhoff79} 
Altenhoff, W.J., Downes, D., Pauls, T., \& Schraml, J. 1979, \aaps, 35, 23

\bibitem[Afflerbach et al. (1996)]{afflerbach96} 
Afflerbach, A., Churchwell, E., Accord, J.M., Hofner, P., Kurtz, S., \&
DePree, C.G. 1996, \apjs, 106, 423

\bibitem[Anderson \& Bania (2008)]{anderson08}
Anderson, L.D. \& Bania, T.M. 2008, \apj, submitted

\bibitem[Araya et al. (2002)] {araya02}
Araya, E., Hofner, P., Churchwell, E., \& Kurtz, S. 2002, \apjs, 138, 63






\bibitem[Benjamin et al. (2003)]{benjamin03}
Benjamin, R. A., et al. 2003, \pasp, 115, 953

\bibitem[Blitz, Fich, \& Stark (1982)]{blitz82}
Blitz, L., Fich, M., \& Stark, A.A. 1982, \apjs, 49, 183

\bibitem[Brand et al. (1984)]{brand84}
Brand, J., van der Bij, M.D.P., de Vries, C.P., Leene, A., Habing, H.J., Israel, F.P., 
de Graauw, T., van de Stadt, H., \& Wouterloot, J.G.A. 1984, \aap, 139, 181

\bibitem[Brand(1986)]{brand86}
Brand, J. 1986, PhD Thesis, Leiden Univ. (Netherlands)

\bibitem[Brogan et al. (2006)]{brogan06}
Brogan, C. L., Gelfand, J. D., Gaensler, B. M., Kassim, N. E., \& Lazio, T. J, 2006, /apj, 639, 25

\bibitem[Bronfman, Nyman, \& May (1996)]{bronfman96}
Bronfman, L., Nyman, L-A. \& May, J. 1996, \aaps, 115, 81

\bibitem[Burton et al. (1975)] {burton75}
Burton, W. B., Gordon, M.A., Bania, T.M., \& Lockman, F.J. 1975,
\apj, 202, 30





\bibitem[Churchwell, Walmsley, \& Cesaroni (1990)]{churchwell90}
Churchwell, E., Walmsley, C.M., \& Cesaroni, R. 1990, \aaps, 83, 119 

\bibitem[Churchwell et al. (2006)]{churchwell06}
Churchwell, E., et al. 2006, \apj, 649, 759

\bibitem[Clemens(1985)]{clemens85}
Clemens, D. P. 1985, \apj, 295, 422

\bibitem[Clemens \& Barvainis(1988)]{clemens88}
Clemens, D. P. \& Barvainis, R. 1988, \apjs, 68, 257

\bibitem[Clark et al. (2003)]{clark03}
Clark, J. S., Egan M. P., Crowther P. A., Mizuno D. R., Larionov V. M., \&
Arkharov A. 2003, \aap, 412, 185


\bibitem[Condon et al. (1998)]{condon98} 
Condon, J. J., Cotton, W. D., Greisen, E. W., Yin, Q. F., Perley, R. A.,
Taylor, G. B., \& Broderick, J. J. 1998, \aj, 115, 1693


\bibitem[Dame, Hartmann \& Thaddeus (2001)] {dame01}
Dame, T. M., Hartmann, D., \& Thaddeus, P. 2001, \apj, 547, 792



\bibitem[Dyson \& Williams (1997)]{dyson97} 
Dyson, J. E., \& Williams, D. A. 1997, The Physics of the Interstellar
Medium (2nd ed.; Bristol: Inst. Physics)


\bibitem[Falgarone \& Phillips (1996)]{falgorone96}
Falgarone E. \& Phillips, T.G. 1996, \apj, 472, 191

\bibitem[Fich, Dahl, \& Treffers (1990)]{fich90}
Fich, L., Dahl, G., \& Treffers, R. 1990, \aj, 99, 622

\bibitem[Flynn et al. (2004)]{flynn04}
Flynn, E. S., Jackson, J. M., Simon, R., Shah, R. U., Bania, T. M. 
\& Wolfire, M. 2004, ASP Conference Series, 317, 44 



\bibitem[Gaensler, Gotthelf, \& Vasisht (1999)]{gaensler99}
Gaensler, B. M., Gotthelf, E. V., \& Vasisht, G. 1999, \apj,, 526, 37




\bibitem[Green (2006)] {green06}
Green, D.A. 2006, A Catalog of Galactic Supernova Remnants (2006
April version), Mullard Radio Astronomy Observatory, Cavendish
Laboratory (Cambridge, United Kingdom)



\bibitem[Helfand et al. (2006)]{helfand06}
Helfand, D.J., Becker, R.H., White, R.L., Fallon, A., and Tuttle,
S. 2006, \aj, 131, 2525



\bibitem[Hollenbach \& Tielens (1997)] {hollenbach97}
Hollenbach, D. J., \& Tielens, A. G. G. M. 1997, \araa, 35, 179




\bibitem[Jackson et al. (2002)] {jackson02}
Jackson, J. M., Bania, T. M., Simon, R., Kolpak, M., Clemens, D. P. 
\& Heyer, M. 2002, \apj, 566, 81 

\bibitem[Jackson et al. (2006)] {jackson06}
Jackson, J. M., Rathborne, J.M., Shah, R.Y., Simon, R., Bania, T.M.,
Clemens, D.P., Chambers, E.T., Johnson, A.M., Dormody, M. \& Lavoie, R. 
2006, \apjs, 163, 145 







\bibitem[Kim \& Koo (2003)]{kim03}
Kim, K. \& Koo, B. 2003, \apj, 596, 362

\bibitem[Kolpak et al. (2003)]{kolpak03}
Kolpak, M.A., Jackson, J.M., Bania, T.M., \& Clemens, D.P. 2003, \apj, 582, 756

\bibitem[Kramer et al. (1998)]{kramer98}
Kramer, C., Stutzki, J., R\"{o}hrig, R., \& Corneliussen, U. 1998, \aap, 329, 249 

\bibitem[Kuchar \& Bania (1994)]{kuchar94}
Kuchar, T.A. \& Bania, T.M. 1994, \apj, 436, 117

\bibitem[Kurtz et al. (1994)] {kurtz94}
Kurtz, S.,  Churchwell, E.,  Wood, D.O.S. \& Myers, P. 1994, \apjs, 91, 659



\bibitem[Kwok, Volk, \& Bidelman (1997)]{kwok97}
Kwok, S., Volk, K., \& Bidelman W.P. \apjs, 1997, 112, 557


\bibitem[Lee et al. (2001)] {lee01}
Lee, Y., Stark, A. A., Kim, H., \& Moon, D. 2001, ApJS, 136, 137

 
\bibitem[Lockman (1989)] {lockman89} 
Lockman, F. J. 1989, \apjs, 71, 469

\bibitem[Lockman, Pisano, \& Howard (1996)] {lockman96} 
Lockman, F. J, Pisano, D. J., \& Howard, G. J., \apj, 472, 173




\bibitem[Paladini et al. (2003)] {paladini03}
Paladini R., Burigana C., Davies R. D., Maino D., Bersanelli M.,
Cappellini B., Platania P., \& Smoot G. 2003, \aap, 397, 213






\bibitem[Rohlfs \& Wilson (1996)] {rohlfs}
Rohlfs, K. \& Wilson, T.L. 1996, Tools of Radio Astronomy, (3rd ed.; Heidelberg: Springer)




\bibitem[Russeil \& Castets (2004)]{russeil04}
Russeil, D. \& Castets, A. 2004, \aap, 417, 107

\bibitem[Sanders et al. (1986)] {sanders86}
Sanders, D.B., Clemens, D.P., Scoville, N.Z., \& Solomon, P.M.
1986, ApJS, 60, 1

\bibitem[Scoville \& Solomon (1975)] {ss75}
Scoville, N. Z. \& Solomon, P. M. 1975, \apjl, 199, L105



\bibitem[Simon et al. (2001)]{simon01}
Simon, R., Jackson, J.M., Clemens, D.P., \& Bania, T.M. 2001, \apj, 551, 747


\bibitem[Sewilo, et al. (2004)]{sewilo04}
Sewilo, M., Churchwell, E., Kurtz, S., Goss, W.M., Hofner, P. 2004, \apj, 605, 285



\bibitem[Stephenson (1992)]{stephenson92}
Stephenson, C.B. 1992, \aj, 103, 263


\bibitem[Stil et al. (2006)]{stil06}
Stil, J. M., et al. 2006, \aj, 132, 1158




\bibitem[Watson et al. (2003)] {watson03}
Watson, C., Araya, E., Sewilo, M., Churchwell, E., Hofner, P., \& Kurtz,
				  S. 2003, \apj, 587, 714

\bibitem[Whiteoak, Otrupcek, \& Rennie (1982)]{whiteoak82}
 Whiteoak, J. B., Otrupcek, R. E., \& Rennie, C. J. 1982, PASAu, 4, 434

\bibitem[Williams, de Geus, \& Blitz (1994)]{williams94}
Williams, J.P., de Geus, E. J., \& Blitz, L. 1994, \apj, 428, 693



\bibitem[Wood \& Churchwell (1989a)] {wc89a}
Wood, D.O.S. \& Churchwell, E. 1989, \apj, 69, 831

\bibitem[Wood \& Churchwell (1989b)] {wc89b}
Wood, D.O.S. \& Churchwell, E. 1989, \apj, 340, 265

\end{thebibliography}
\end{document}